\documentclass[aps,superscriptaddress,nofootinbib,11pt]{revtex4-1} 
\pdfoutput=1

\usepackage{dcolumn}    
\usepackage{pstricks}
\usepackage{color}
\usepackage{graphicx}
\usepackage{amsmath}
\usepackage{mathtools}
\usepackage[T1]{fontenc}
\usepackage[utf8]{inputenc}
\usepackage{amssymb}
\usepackage[colorlinks=true,allcolors=darkred, pdfborder={0 0 0}]{hyperref}
\usepackage[caption=false]{subfig}
\usepackage{longtable}
\usepackage{needspace}
\usepackage{pifont}
\newcommand{\cmark}{\ding{51}}%
\newcommand{\xmark}{\ding{55}}%

\usepackage{booktabs}

\allowdisplaybreaks

\def\MGUT{M_\mathrm{GUT}}

\def\MI{M_\mathrm{I}}
\def\MZ{M_Z}
\def\SO10{\text{SO}(10)}
\newcommand{\G}[1]{\mathcal{G}_\text{#1}}
\def\SU{\,\text{SU}}
\def\GeV{\,\mathrm{GeV}}

\newcommand{\rep}[1]{\mathbf{#1}}
\newcommand{\repb}[1]{\mathbf{\overline{#1}}}

\definecolor{darkred}{rgb}{0.6,0,0}

\begin{document}

\title{Threshold effects in SO(10) models with one intermediate breaking scale}

\author{Davide Meloni}
\email{davide.meloni@uniroma3.it}
\affiliation{Dipartimento di Matematica e Fisica,
	Università di Roma Tre,
	Via della Vasca Navale 84, 
	00146 Rome, 
	Italy}

\author{Tommy Ohlsson}
\email{tohlsson@kth.se}
\affiliation{Department of Physics,
	School of Engineering Sciences,
	KTH Royal Institute of Technology,
	AlbaNova University Center,
	Roslagstullsbacken 21,
	SE--106 91 Stockholm,
	Sweden}
\affiliation{The Oskar Klein Centre,
	AlbaNova University Center,
	Roslagstullsbacken 21,
	SE--106 91 Stockholm,
	Sweden}
\affiliation{University of Iceland, 
	Science Institute, 
	Dunhaga 3, 
	IS--107 Reykjavik, 
	Iceland}

\author{Marcus Pernow}
\email{pernow@kth.se}
\affiliation{Department of Physics,
	School of Engineering Sciences,
	KTH Royal Institute of Technology,
	AlbaNova University Center,
	Roslagstullsbacken 21,
	SE--106 91 Stockholm,
	Sweden}
\affiliation{The Oskar Klein Centre,
	AlbaNova University Center,
	Roslagstullsbacken 21,
	SE--106 91 Stockholm,
	Sweden}

\begin{abstract}
Despite the successes of the Standard Model of particle physics, it is known to suffer from a number of deficiencies. Several of these can be addressed within non-supersymmetric theories of grand unification based on $\SO10$. However, achieving gauge coupling unification in such theories is known to require additional physics below the unification scale, such as symmetry breaking in multiple steps. Many such models are disfavored due to bounds on the proton lifetime. Corrections arising from threshold effects can, however, modify these conclusions. We analyze all seven relevant breaking chains with one intermediate symmetry breaking scale, assuming the ``survival hypothesis'' for the scalar masses. Two are allowed by proton lifetime and two are disfavored by a failure to unify the gauge couplings. The remaining three unify at a too low scale, but can be salvaged by various amounts of threshold corrections. We parametrize this and thereby rank the models by the size of the threshold corrections required to save them.
\end{abstract}

\maketitle

\section{Introduction}
Grand unified theories (GUTs) in general~\cite{Georgi:1974sy}, and in particular models based on the $\SO10$ gauge symmetry~\cite{Fritzsch:1974nn}, are popular extensions of the Standard Model (SM) of particle physics. They can provide solutions to a number of open questions in the SM, such as the nature of charge quantization, anomaly cancellation, and the existence of three separate gauge groups~\cite{Langacker:1980js}. Of a more phenomenological nature, $\SO10$ models naturally account for the generation of small neutrino masses through the type I~\cite{Minkowski:1977sc,GellMann:1980vs,Mohapatra:1979ia,Yanagida:1979,Schechter:1980gr} or type II~\cite{Magg:1980ut,Lazarides:1980nt,Mohapatra:1980yp} seesaw mechanisms. 

A prerequisite of grand unification is that the evolution of the SM gauge couplings with energy scale, governed by the renormalization group equations (RGEs), must be such that they unify. It is well-known that the gauge couplings do not unify in non-supersymmetric (non-SUSY) models unless intermediate symmetry breaking scales or fields with intermediate-scale masses are added, but that successful gauge coupling unification can be achieved in the minimal supersymmetric SM~\cite{Amaldi:1991cn,Giunti:1991ta,Langacker:1991an}. On the other hand, since the $\SO10$ group is of rank five, which is one larger than the SM gauge group, the symmetry breaking may occur in multiple steps~\cite{Georgi:1979ga,delAguila:1980qag,Rizzo:1981su,Rizzo:1981dm,Buccella:1980qb,Rajpoot:1980xy,Yasue:1980qj,Yasue:1980fy,Anastaze:1983zk,Gipson:1984aj,Babu:1984mz,Chang:1984qr,Bertolini:2009qj,Bertolini:2009es}. This modifies the evolution of the gauge couplings in a way that can allow their unification even in non-SUSY models.

Currently, the most constraining experimental prediction of GUTs is the instability of protons. The additional leptoquark scalar and gauge bosons that reside at the scale of unification $\MGUT$ in general mediate proton decay. This, together with the non-observation of proton decay, places a lower bound on the value of $\MGUT$. In turn, this can disfavor some of the possible intermediate gauge groups since they predict a value of $\MGUT$ that is too low~\cite{Deshpande:1992au}. 

Threshold corrections~\cite{Weinberg:1980wa,Hall:1980kf} are loop level corrections arising from fields lying at and around the scale of symmetry breaking that modify the matching conditions of the gauge couplings of the models above and below the energy scale of symmetry breaking. This can in turn modify the value of $\MGUT$  and thereby save some of the models that were previously disfavored~\cite{Parida:1989an,Rani:1993pp,Mohapatra:1992dx,Lee:1994vp,Bertolini:2009qj,Babu:2015bna,Parida:2016hln,Chakraborty:2019uxk,Chakrabortty:2019fov}. Furthermore, since threshold corrections modify the matching conditions of the gauge couplings, they can allow for unification in models where the gauge couplings do not unify~\cite{Lavoura:1993su,Ellis:2015jwa,Schwichtenberg:2018cka}. The threshold corrections also impact the intermediate scale $\MI$, which is relevant, for example, for neutrino masses.

In this work, we consider the direct breaking of $\SO10$ to the SM as well as all relevant models with one intermediate symmetry breaking scale. For the model with direct breaking to the SM, we study how the threshold corrections may allow for unification without the addition of intermediate symmetries. For the models with an intermediate symmetry breaking step in which unification is achieved, we investigate how the threshold corrections affect $\MGUT$ and $\MI$. Thereby, we quantify how large threshold corrections are required in order to save the models that \textit{a priori} predict a proton lifetime that is too short. The renormalization group (RG) running is performed at two-loop level, with the one-loop level result given for comparison. We do not take into account any restrictions on $\MI$ from neutrino masses in order to refrain from making too many assumptions about the models. This work differs significantly from recent works dealing with threshold correction~\cite{Bertolini:2013vta,Babu:2015bna,Schwichtenberg:2018cka,Chakrabortty:2019fov} in that we consider all possible breaking chains with at most one intermediate step and treat all models in the same way.  This, together with our comprehensive numerical analysis of the effect of threshold corrections, allows a simple and quantitative comparison between the plausibility of the different breaking chains.

In Sec.~\ref{sec:models}, we discuss the models that are analyzed in this work and present the solutions to the RGEs. We comment on which models achieve gauge coupling unification and the prediction for the proton lifetime in each of the models. Then, in Sec.~\ref{sec:thresholds}, we describe the computation of threshold corrections. In Sec.~\ref{sec:results}, we present the results of the threshold corrections for the different models. Finally, in Sec.~\ref{sec:conclusion}, we summarize our findings and conclude.

\section{Models}\label{sec:models}
In this section, we discuss the eight models that we investigate in this work. Furthermore, we give some details on the particle content that is involved in each model and comment on the RG running.

The most minimal non-SUSY $\SO10$-based model is one in which the gauge symmetry is broken directly to the SM. Following this logic, the next-to-minimal breaking chains are those with one intermediate gauge symmetry. The possible intermediate breaking chains may be seen, for example, in Refs.~\cite{Chang:1984qr,Ferrari:2018rey}. Here, we consider the direct breaking of $\SO10$ to the SM as well as all models with one intermediate symmetry breaking scale with at least two group factors in the intermediate symmetry. The reason that we require at least two group factors is that if there is only one, \textit{e.g.} $\SU(5)$, then the intermediate symmetry does not impact the possibility of gauge coupling unification, unless lighter fields are added as in \textit{e.g.} Ref.~\cite{Boucenna:2017fna}.

In all models, the fermionic particle content consists of three generations of the spinorial $\rep{16}_F$. In order to accommodate realistic fermion mass and mixing parameters, the scalar sector contains a complexified $\rep{10}_H$ representation~\cite{Panagiotakopoulos:1985pt,Babu:1992ia,Bajc:2005zf} and a $\repb{126}_H$ representation. In order to retain some predictivity of the Yukawa sector of $\SO10$, we impose a Peccei--Quinn (PQ) symmetry~\cite{Peccei:1977hh,Peccei:1977ur}, which forbids one of the two independent couplings between the fermions and the $\rep{10}_H$~\cite{Holman:1982tb}. This is not necessary for $\SO10$ model building, but is often invoked in realistic models. In general, the breaking of the PQ symmetry leads to the axion domain wall problem, in which domain walls between different vacua are generated and dominate the Universe~\cite{Sikivie:1982qv}. We do not analyze this problem in detail in the models investigated in this work, but note that there exist solutions, such as the Lazarides--Shafi mechanism~\cite{Lazarides:1982tw}.

We assume that below the intermediate symmetry breaking scale $\MI$, only the SM particle content survives and all other multiplets have masses around either $\MGUT$ or $\MI$. This is in accordance with the ``survival hypothesis'', namely that scalars acquire the largest possible mass that is compatible with the symmetry breaking~\cite{Georgi:1979md,delAguila:1980qag,Mohapatra:1982aq,Dimopoulos:1984ha}.

To one-loop order in perturbation theory, gauge couplings evolve from one scale $M$ to another scale $\mu$ according to
\begin{equation}
\alpha_i^{-1} (\mu) = \alpha_i^{-1} (M) - \frac{a_i}{2\pi}\ln\left(\frac{\mu}{M} \right),
\end{equation}
where the index $i$ denotes the group to which the gauge coupling corresponds. The coefficient $a_i$, known as the $\beta$ coefficient, is determined by the particle content that exists in the relevant energy regime. It is given by~\cite{Jones:1981we,Machacek:1983tz}
\begin{equation}
a_i = -\frac{11}{3}C_2(G_i) +\frac{4}{3}\kappa_F S_2(F_i) + \frac{1}{3} \kappa_S S_2(S_i),
\end{equation}
where $C_2(r)$ is the quadratic Casimir and $S_2(r)$ [sometimes also denoted $C(r)$] is the Dynkin index of the representation $r$, related to the quadratic Casimir by
\begin{equation}
S_2(r) = \frac{d(r)}{d(G)}C_2(r),
\end{equation}
where $G$ refers to the adjoint representation and $d(r)$ denotes the dimension of representation $r$. Furthermore, $F_i$ is the representation that the fermions belong to and $S_i$ is the representation of the scalars. The coefficient $\kappa_F$ is $1$ for Dirac fermions and $1/2$ for Weyl fermions and $\kappa_S$ is $1$ for complex scalars and $1/2$ for real scalars.

To two-loop order in perturbation theory, the gauge couplings obey the differential equation
\begin{equation}
\frac{\text{d}\,\alpha_i^{-1}(\mu)}{\text{d}\ln\mu} = -\frac{a_i}{2\pi} - \sum_j \frac{b_{ij}}{8\pi^2 \alpha_j^{-1}(\mu)},
\end{equation}
where the two-loop coefficients are given by~\cite{Jones:1981we,Machacek:1983tz}
\begin{equation}\label{eq:2loop}
b_{ij} = -\frac{34}{3}\left[ C_2(G_i)\right]^2 \delta_{ij} + \kappa_F\left[4C_2(F_j) + \frac{20}{3} C_2(G_i)\delta_{ij} \right]S_2(F_i) + \kappa_S \left[4C_2(S_j) + \frac{2}{3} C_2(G_i)\delta_{ij} \right]S_2(S_i).
\end{equation}
There is also a contribution from the Yukawa coupling to the two-loop $\beta$ function above. However, since the RG running of the Yukawa couplings is somewhat model-dependent, we neglect that term in Eq.~\eqref{eq:2loop}. The $\beta$ coefficients $a_i$ and $b_{ij}$ for the models considered are listed in Tab.~\ref{tab:beta} in App.~\ref{app:beta}.

Given the values of the gauge couplings at the electroweak scale $\MZ \simeq 91.1876\GeV$~\cite{Tanabashi:2018oca},
\begin{equation}
\left(\alpha^{-1}_3(\MZ),\, \alpha^{-1}_2(\MZ),\, \alpha^{-1}_1(\MZ)\right) = (8.50, \,29.6,\, 59.0),
\end{equation}
the system of RGEs can be solved.\footnote{Note that we use the central values and neglect their uncertainties. Taking into account the uncertainties, the largest of which is about $0.8~\%$ on $\alpha_3(\MZ)$~\cite{Tanabashi:2018oca}, would impact the computed scales $\MI$ and $\MGUT$ by less than 5~\%, which does not affect our conclusions.} Depending on the $\beta$ functions, precise gauge coupling unification may be obtained. If it is possible, then that model is an allowed model for grand unification. 

The relevant experimental prediction of grand unification related to the scale of unification is proton decay. From the scale of grand unification and the coupling strength $g_\text{GUT}$ at that scale, the proton lifetime in the most constraining channel can be computed as~\cite{Nath:2006ut,Babu:2010ej}
\begin{equation}
\Gamma(p\rightarrow e^+\pi^0 ) \simeq \frac{m_p}{64\pi f_\pi^2} \frac{g_\text{GUT}^4}{\MGUT^4}A_L^2 \alpha_H^2 F_q,
\end{equation}
where $f_\pi\simeq 139\,\text{MeV}$ is the pion decay constant, $A_L\simeq 2.726$ is a renormalization factor, $\alpha_H\simeq 0.012\GeV ^3$ is the hadronic matrix element, and $F_q\simeq 7.6$ is a quark-mixing factor. With these numerical factors, the proton lifetime in this channel can be estimated by
\begin{equation}
\tau(p\rightarrow e^+\pi^0)\simeq (7.5\times 10^{35}\, \text{yr}) \left(\frac{\MGUT}{10^{16}\GeV}\right)^4 \left(\frac{0.03}{\alpha_\text{GUT}}\right)^2.
\end{equation}
Since proton decay has not been experimentally observed, there is a lower bound on the lifetime of the proton. The most constraining one is from Super-Kamiokande~\cite{Mine:2016mxy,Miura:2016krn,Abe:2014mwa} with the bound $\tau(p\rightarrow e^+\pi^0)>1.67\times 10^{34}\,\text{yr}$ at $90~\%$ confidence level. Any model must be able to accommodate a proton lifetime longer than the experimental bound.

In what follows, we employ the conventions that gauge couplings, $\beta$ coefficients, and representations of fields appear in the order in which the gauge group is written. For example, in $\SU(3)_\text{C}\times \SU(2)_\text{L}\times\text{U}(1)_Y$, the first entry corresponds to $\SU(3)_\text{C}$, the second to $\SU(2)_\text{L}$, and the third to $\text{U}(1)_Y$. For representations of fields, Abelian charges are always listed as subscripts.

\subsection{No Intermediate Symmetry}
The direct breaking of the $\SO10$ symmetry to the SM gauge group $\G{321} = \SU(3)_\text{C}\times \SU(2)_\text{L}\times\text{U}(1)_Y$ can be achieved with a $\rep{144}_H$ taking a vacuum expectation value (vev) in the appropriate direction~\cite{Babu:2005gx}. We then assume that all multiplets from within the $\rep{144}_H$ have masses at $\MGUT$. Assigning a non-zero PQ charge to the $\rep{144}_H$ allows it to also break the PQ symmetry at $\MGUT$ since the vev of a charged component of the field results in spontaneous symmetry breaking. Further, we assume that from the $\rep{10}_H$ and the $\repb{126}_H$, only one combination of the $\SU(2)_\text{L}$ doublets survives below $\MGUT$. This is the SM Higgs doublet~\cite{Bajc:2005zf}, such that the SM particle content is recovered below $\MGUT$. The other fields that are not part of the SM field content reside at $\MGUT$. These are listed in Tab.~\ref{tab:SM} in App.~\ref{app:fields}.

From the particle content described above, the $\beta$ coefficients $a_i$ and $b_{ij}$ may be computed. They are listed in Tab.~\ref{tab:beta} in App.~\ref{app:beta}. The resulting evolution of the SM gauge couplings is shown in Fig.~\ref{fig:SM}. As is well known, the gauge couplings fail to unify.

\subsection{$\SU\text{(4)}\times\SU\text{(2)}\times\SU\text{(2)}$}\label{sec:422}
A popular model of intermediate symmetry in the breaking of $\SO10$ is the Pati--Salam (PS) model, based on the gauge symmetry $\G{422} = \SU(4)_\text{C}\times \SU(2)_\text{L}\times \SU(2)_\text{R}$~\cite{Pati:1974yy}. It is a maximal subgroup of $\SO10$ and contains the SM gauge group $\G{321}$ as a subgroup. The fermions are embedded in this model as $(\rep{4},\rep{2},\rep{1})\oplus (\repb{4},\rep{1},\rep{2})$.

Models based on a Pati--Salam intermediate symmetry have been investigated extensively in the literature~\cite{Lazarides:1980nt,Babu:1992ia,Lavoura:1993vz,Lee:1994vp,Buccella:2012kc,Meloni:2014rga,Meloni:2016rnt,Babu:2016bmy,Ohlsson:2018qpt}. In this work, we follow the model described in Ref.~\cite{Altarelli:2013aqa}, in which the $\SO10$ symmetry is broken by a vev in the $\rep{210}_H$. To break the PS symmetry as well as the PQ symmetry down to $\G{321}$, we employ a vev in the $\repb{126}_H$ together with a complex $\rep{45}_H$ with a non-zero PQ charge.

The reason that two separate vevs are needed even though the $\repb{126}_H$ has a non-zero PQ charge is to break the linear combination of PQ, $B-L$, and $T_{3,\text{R}}$ which is otherwise left invariant~\cite{Holman:1982tb,Mohapatra:1982tc,Holman:1982tb,Bajc:2005zf}. Although the minimal choice of an extra representation for the breaking of the PQ symmetry could be considered to be a singlet, we will not use this. The reason is that singlets have mass terms that are unprotected by any symmetry and this choice would therefore introduce unnecessary fine-tuning. Thus a $\rep{45}_H$ is introduced.

Between $\MGUT$ and $\MI$, the scalar fields are $(\rep{1},\rep{2},\rep{2})$ from the $\rep{10}_H$, $(\rep{15},\rep{2},\rep{2})\oplus(\rep{10},\rep{1},\rep{3})$ from the $\repb{126}_H$, and $(\rep{1},\rep{1},\rep{3})$ from the $\rep{45}_H$. From these, the $\beta$ coefficients, listed in App.~\ref{app:beta}, can be computed. The fields that lie at $\MGUT$ and $\MI$ are given in Tab.~\ref{tab:PS} in App.~\ref{app:fields}.

To compute the RG running in this model, one also requires the matching conditions between the SM and the PS models. The gauge couplings of the model based on $\G{422}$ at $\MI$ are derived from the gauge couplings of the SM by
\begin{align}
\alpha_4^{-1}(\MI) &= \alpha_3^{-1}(\MI),  \label{eq:match3}\\
\alpha_\text{2L}^{-1}(\MI) &= \alpha_2^{-1}(\MI), \label{eq:match2}\\
\alpha_\text{2R}^{-1}(\MI) &= \frac{5}{3} \alpha_Y^{-1}(\MI) - \frac{2}{3}\alpha_3^{-1}(\MI) .\label{eq:matchY}
\end{align}
The resulting RG evolution is shown in Fig.~\ref{fig:PS}. At two-loop level, the solutions for the scales gives an intermediate scale of $\MI \approx 2.64\times 10^{9}\GeV$ and a unification scale of $\MGUT \approx 3.72\times 10^{16}\GeV$, corresponding to a proton lifetime of $\tau_p\approx 1.2\times 10^{38}\,\text{yr}$.\footnote{In comparison, at one-loop level, the scales are $\MI \approx 1.28\times 10^{11}\GeV$ and $\MGUT \approx 1.96\times 10^{16}\GeV$, corresponding to a proton lifetime of $\tau_p\approx 1.3\times 10^{37}\,\text{yr}$.} Hence, it is clear that the PS model is allowed by the proton lifetime limit.

\subsection{$\SU\text{(4)}\times\SU\text{(2)}\times\SU\text{(2)}\times \text{D}$}
This model is based on a PS gauge symmetry with an additional left-right $D$ parity which acts on the fields such that $(\rep{r_4},\rep{r_L},\rep{r_R})\rightarrow(\repb{r_4},\rep{r_R},\rep{r_L})$~\cite{Mohapatra:1974hk,Mohapatra:1974gc,Senjanovic:1975rk,Kibble:1982ae}. For a previous analysis of a similar model see \textit{e.g.} Ref.~\cite{Lee:1994vp,Babu:2015bna}.

In order to preserve $D$ parity when breaking the $\SO10$ symmetry, a $\rep{54}_H$ can be used. The breaking of the $\G{422D}$ symmetry can be achieved using the $\repb{126}_H$. A $\rep{45}_H$ is used to also break the PQ symmetry, as described in Sec.~\ref{sec:422}. 

Between $\MGUT$ and $\MI$, we have $(\rep{1},\rep{2},\rep{2})$ from the $\rep{10}_H$ and $(\rep{15},\rep{2},\rep{2})\oplus (\rep{10},\rep{1},\rep{3})\oplus (\repb{10},\rep{3},\rep{1})$ from the $\repb{126}_H$. Note that the latter is needed only to preserve $D$ parity of the model. From the $\rep{45}_H$, we have $(\rep{1},\rep{1},\rep{3})\oplus (\rep{1},\rep{3},\rep{1})$, where, again, the latter representation only serves to conserve $D$ parity. Note that the difference between this model and the one in Ref.~\cite{Babu:2015bna} is that they do not include the $\rep{45}_H$ and that they place $(\rep{6},\rep{1},\rep{1})$ from the $\repb{126}_H$ at $\MI$ due to considerations of the scalar potential, which are beyond the scope of this work. The fields that lie at $\MGUT$ and $\MI$ are given in Tab.~\ref{tab:PSD} in App.~\ref{app:fields}.

With the particle content described, we can compute the $\beta$ coefficients which are given in App.~\ref{app:beta}. The matching conditions are the same as for the PS model, namely Eqs.~\eqref{eq:match2}--\eqref{eq:matchY}, with the additional constraint that $\alpha_\text{2L}^{-1}=\alpha_\text{2R}^{-1}$ due to $D$ parity. With these, one can calculate the RG evolution of the gauge couplings and the required intermediate scale for their unification, as shown in Fig.~\ref{fig:PSD}. At two-loop level, the intermediate scale is $\MI\approx 4.34\times 10^{13}\GeV$ and the unification scale is $\MGUT\approx 7.45\times 10^{14}\GeV$, giving a proton lifetime of $\tau_p\approx 3\times 10^{31}\,\text{yr}$, below the experimental lower limit, as previously noted in \textit{e.g.}~\cite{Deshpande:1992au}.\footnote{At one-loop level, the scales are $\MI\approx 5.00\times 10^{13}\GeV$ and $\MGUT\approx 1.40\times 10^{15}\GeV$, resulting in a proton lifetime of $\tau_p\approx 3.7\times 10^{32}\,\text{yr}$, which is also too short.}

\subsection{$\SU\text{(4)}\times\SU\text{(2)}\times\text{U(1)}$}
The gauge group $\G{421}=\SU(4)_\text{C}\times\SU(2)_\text{L}\times\text{U}(1)_\text{R}$ is a subgroup of the PS gauge group, but it may be reached directly by breaking the $\SO10$ symmetry. This is possible by assigning a vev to the appropriate direction of the $\rep{45}_H$. Models based on this gauge group have been previously analyzed in \textit{e.g.} Refs.~\cite{Holman:1982tb,Parida:1991sj,Rani:1993pp}. The breaking of $\G{421}$ down to $\G{321}$ can then be done using a vev of $(\repb{10},\rep{1})_1$ from the $\repb{126}_H$. Since now the $\rep{45}_H$, which carries a PQ charge, is used to break the symmetry at $\MGUT$, the PQ symmetry will also be broken at that scale. Contrary to the  previously discussed models, we do not need to include two separate vevs to break the remaining linear combination of charges, since now both $B-L$ and $\text{R}$ remain unbroken at $\MGUT$.

To compute the $\beta$ coefficients between $\MI$ and $\MGUT$, we first need to list the various fields that are present between those two scales. For the fermions, we have $(\rep{4},\rep{2})_0\oplus (\repb{4},\rep{1})_{1/2}\oplus (\repb{4},\rep{1})_{-1/2}$. The scalars are $(\rep{1},\rep{2})_{-1/2}$ from the $\rep{10}_H$ and $(\rep{15},\rep{2})_{-1/2}\oplus (\rep{10},\rep{1})_{-1}$ from the $\repb{126}_H$. The resulting $\beta$ coefficients are found in App.~\ref{app:beta} and the resulting gauge coupling unification can be seen in Fig.~\ref{fig:421}. The matching conditions between $\G{321}$ and $\G{421}$ are identical to those given in Eqs.~\eqref{eq:match2}--\eqref{eq:matchY}, with the replacement $\alpha_\text{2R}^{-1}(\MI) \rightarrow \alpha_\text{1R}^{-1}(\MI)$. The fields that lie at $\MGUT$ and $\MI$ are given in Tab.~\ref{tab:421} in App.~\ref{app:fields}.

At two-loop level, the resulting scales are $\MI\approx 1.57\times 10^{11}\GeV$ and $\MGUT \approx 2.69\times 10^{14}\GeV$, giving a proton lifetime of $\tau_p\approx 5.8\times 10^{29}\,\text{yr}$.\footnote{At one-loop level, the scales are $\MI\approx 1.35\times 10^{11}\GeV$ and $\MGUT\approx 4.60\times 10^{14}\GeV$, corresponding to a proton lifetime of $\tau_p\approx 5.1\times 10^{30}\,\text{yr}$.} Thus, as noted previously in the literature~\cite{Deshpande:1992au}, this model is ruled out by proton decay bounds.

\subsection{$\SU\text{(3)}\times\SU\text{(2)}\times\SU\text{(2)}\times \text{U(1)}$}\label{sec:3221}
Another subgroup of the PS gauge group is $\G{3221}=\SU(3)_\text{C}\times\SU(2)_\text{L}\times\SU(2)_\text{R}\times \text{U}(1)_{B-L}$, as studied in \textit{e.g.} Refs.~\cite{Senjanovic:1978ev,Holman:1982tb,Chang:1983fu,Chang:1984uy,Lee:1994vp,Arbelaez:2013nga,Patra:2015bga,Bandyopadhyay:2015fka,Chakrabortty:2017mgi,Lazarides:2020frf}. This may be reached by direct breaking of the $\SO10$ symmetry by \textit{e.g.} a vev in the $\rep{45}_H$. Similar to the $\G{421}$ model, the PQ symmetry is broken at $\MGUT$ by the $\rep{45}_H$ so that only one vev is required to break the symmetry at $\MI$. The breaking of $\G{3221}$ down to $\G{321}$ can be achieved with $(\rep{1},\rep{1},\rep{3})_2$ from the $\repb{126}_H$.

Computing the $\beta$-functions between $\MI$ and $\MGUT$, we first note that the fermions are embedded as $(\rep{3},\rep{2},\rep{1})_{1/3}\oplus (\rep{1},\rep{2},\rep{1})_{-1}\oplus (\repb{3},\rep{1},\rep{2})_{-1/3}\oplus (\rep{1},\rep{1},\rep{2})_1$. The scalars that are between those two scales are $(\rep{1},\rep{2},\rep{2})_0$ from the $\rep{10}_H$ as well as $(\rep{1},\rep{2},\rep{2})_0\oplus (\rep{1},\rep{1},\rep{3})_{-2}$ from the $\repb{126}_H$. Based on these fields, the $\beta$ coefficients can be found in App.~\ref{app:beta}. The fields that lie at $\MGUT$ and $\MI$ are given in Tab.~\ref{tab:3221} in App.~\ref{app:fields}.

In this model, the matching condition at $\MI$ is more involved than in the above-discussed models due to the fact that the $B-L$ needs to be appropriately normalized. Before normalization, the hypercharge $Y$ may be expressed as
\begin{equation}
Y = \frac{B-L}{2} - T_{3R}.
\end{equation}
In order to normalize these charges, the hypercharge $Y$ is multiplied by the GUT normalization factor of $\sqrt{3/5}$ and the $B-L$ charge is multiplied by $\sqrt{3/8}$. From this, one can derive the matching conditions of the appropriately normalized gauge couplings, namely
\begin{equation}
\alpha_Y^{-1} = \frac{2}{5} \alpha_{B-L}^{-1} + \frac{3}{5}\alpha_{R}^{-1}.
\end{equation}
In order to invert this relation, we face the issue that we are matching three gauge couplings to four. Thus, we introduce the parameter $x$ such that $\alpha_{B-L}^{-1}(\MI) = x\alpha_\text{R}^{-1}(\MI)$. This parameter is then solved for together with the scales $\MGUT$ and $\MI$ such that gauge coupling unification is achieved. The resulting matching conditions are
\begin{align}
\alpha_3^{-1}(\MI) &= \alpha_3^{-1}(\MI),\\
\alpha_{2\text{L}}^{-1}(\MI) &= \alpha_2^{-1}(\MI), \\
\alpha_\text{2R}^{-1}(\MI) &= \left(\frac{2}{5} x + \frac{3}{5}\right)^{-1} \alpha_Y^{-1}(\MI), \\
\alpha_{B-L}^{-1}(\MI) &= x\left(\frac{2}{5} x + \frac{3}{5}\right)^{-1} \alpha_Y^{-1}(\MI).
\end{align}
The RG running is shown in Fig.~\ref{fig:3221}. 

Solving for the scales that result in unification, we obtain at two-loop level $\MI \approx 1.57\times 10^{10}\,\text{GeV}$ and $\MGUT \approx 5.18\times 10^{15}\,\text{GeV}$ with the parameter $x\approx1.38$, resulting in a proton lifetime of $\tau_p\approx 9.4\times 10^{34}\,\text{yr}$.\footnote{At one-loop level, the result is $\MI \approx 6.59\times 10^{9}\,\text{GeV}$ and $\MGUT \approx 1.39\times 10^{16}\,\text{GeV}$ with the parameter $x\approx1.43$, resulting in $\tau_p\approx 5.0\times 10^{36}\,\text{yr}$.} Therefore, this model is allowed by proton lifetime considerations.

\subsection{$\SU\text{(3)}\times\SU\text{(2)}\times\SU\text{(2)}\times \text{U(1)}\times \text{D}$}
A similar model to the one in Sec.~\ref{sec:3221} but with a surviving $D$ parity may be constructed. In this case, the $\SO10$ symmetry is broken down to the group $\G{3221D}=\SU(3)_\text{C}\times\SU(2)_\text{L}\times\SU(2)_\text{R}\times \text{U}(1)_{B-L}\times D$ by a vev in the $\rep{210}_H$, which conserves the $D$ parity. Such models have been previously studied in \textit{e.g.} Ref.~\cite{Mohapatra:1974gc,Senjanovic:1975rk,Buccella:1987hq,Lee:1994vp,Patra:2015bga,Bandyopadhyay:2015fka}. Then, $\G{3221D}$ is broken to $\G{321}$ using $(\rep{1},\rep{1},\rep{3})_2$ from the $\repb{126}_H$. The PQ symmetry is broken at $\MI$ by $(\rep{1},\rep{1},\rep{3})_0$ from the $\rep{45}_H$.

The fermion embedding is the same as in Sec.~\ref{sec:3221}. For the scalars, between $\MGUT$ and $\MI$, there is $(\rep{1},\rep{2},\rep{2})_0$ from the $\rep{10}_H$. From the $\repb{126}_H$, we have $(\rep{1},\rep{1},\rep{2})_0$ for the SM Higgs and $(\rep{1},\rep{1},\rep{3})_{-2}$ for the symmetry breaking. To conserve $D$ parity, we also need to have $(\rep{1},\rep{3},\rep{1})_{2}$. From the $\rep{45}_H$, we have $(\rep{1},\rep{1},\rep{3})_0$ which is used to break the PQ symmetry as well as $(\rep{1},\rep{3},\rep{1})_0$ in order to conserve $D$ parity. The fields that lie at $\MGUT$ and $\MI$ are given in Tab.~\ref{tab:3221D} in App.~\ref{app:fields}.

From these fields, the $\beta$ coefficients may be calculated and are given in App.~\ref{app:beta}. Although this model has a similar gauge structure to the one in Sec.~\ref{sec:3221}, the matching conditions become somewhat simpler due to the requirement that $\alpha_{2\text{L}}^{-1}=\alpha_{2\text{R}}^{-1}$. This removes the extra freedom introduced by the parameter $x$ above and the matching conditions simply read
\begin{align}
\alpha_3^{-1}(\MI) &= \alpha_3^{-1}(\MI),\\
\alpha_{2\text{L}}^{-1}(\MI) &= \alpha_2^{-1}(\MI), \\
\alpha_\text{2R}^{-1}(\MI) &= \alpha_2^{-1}(\MI), \\
\alpha_{B-L}^{-1}(\MI) &= \frac{5}{2}\alpha_Y^{-1}(\MI) + \frac{3}{2} \alpha_2^{-1}(\MI).
\end{align}
The resulting gauge coupling running is shown in Fig.~\ref{fig:3221D}. 

The scales that result in gauge coupling unification with at two-loop level are $\MI\approx 3.13\times 10^{11}\GeV$ and $\MGUT \approx 6.31\times 10^{14}\GeV$, resulting in a proton lifetime of $\tau_p\approx 1.9\times 10^{31}\,\text{yr}$.\footnote{At one-loop level, the scales are $\MI\approx 1.69\times 10^{11}\GeV$ and $\MGUT\approx 1.29\times 10^{15}\GeV$, with a proton lifetime of $\tau_p\approx 3.3\times 10^{32}\,\text{yr}$.} This model is therefore disfavored due to its prediction of the proton lifetime.

\subsection{$\SU\text{(3)}\times\SU\text{(2)}\times\text{U}\text{(1)}\times \text{U(1)}$}
A subgroup of the $\G{3221}$ group is $\G{3211} = \SU(3)_\text{C}\times\SU(2)_\text{L}\times\text{U}(1)_\text{R}\times \text{U}(1)_{B-L}$, which may also be reached directly by breaking the $\SO10$ symmetry. In order to do so, the singlet inside the $\rep{210}_H$ which also breaks $\G{421}$ and $\G{3221}$ must take a vev. For models based on this gauge group, see \textit{e.g.} Refs.~\cite{Chang:1984uy,Parida:1989an,Parida:1991sj}. The breaking of $\G{3211}$ down to $\G{321}$ can then be achieved using $(\rep{1},\rep{1})_{1,2}$ in the $\repb{126}_H$. In order to also break the remaining combination of the Abelian charges and the PQ symmetry, a vev must be taken by one of the singlets in the $\rep{45}_H$.

The fermions are embedded as $(\rep{3},\rep{2})_{0,1/3}\oplus (\rep{1},\rep{2})_{0,-1}\oplus (\repb{3},\rep{1})_{-1/2,-1/3}\oplus (\repb{3},\rep{1})_{1/2,-1/3}\oplus (\rep{1},\rep{1})_{1/2,1}\oplus (\rep{1},\rep{1})_{-1/2,1}$. The scalars that contribute to the RG running of the gauge couplings are $(\rep{1},\rep{2})_{-1/2, 0}$ from the $\rep{10}_H$ and $(\rep{1},\rep{2})_{-1/2,0}$ as well as $(\rep{1},\rep{1})_{1,2}$ from the $\repb{126}_H$. The field from the $\rep{45}_H$ does not contribute since it is a singlet. 

From this, one can compute the $\beta$ coefficients, given in App.~\ref{app:beta}, as well as the RG running using the same matching conditions as in Sec.~\ref{sec:3221}, replacing $\alpha_\text{2R}^{-1}$ by $\alpha_\text{1R}^{-1}$. Note that in general the RG running is affected by kinetic mixing between the two Abelian gauge factors~\cite{Holdom:1985ag,delAguila:1988jz,Luo:2002iq,Bertolini:2009qj,Lyonnet:2016qyu}. The result is shown in Fig.~\ref{fig:3211}, from which it is clear that unification is not achieved in this model. The reason is that the slopes of the two lines corresponding to the Abelian gauge couplings are too similar, meaning that they do not converge.\footnote{The effect of kinetic mixing is expected to be on the level of a few percent~\cite{Lyonnet:2016qyu} and will therefore not be large enough to allow for gauge coupling unification. This motivates our choice to neglect it in the present work.} Therefore, this model is disfavored on that ground and no prediction of the proton lifetime can be made.

\subsection{$\SU\text{(5)}\times\text{U(1)}$}
The final model considered is the breaking of $\SO10$ to a model of the $\SU(5)$ type. The reason that $\SU(5)$ is not considered on its own is that an intermediate symmetry that is a simple group does not help in achieving gauge coupling unification and instead changes the problem to requiring unification of the three gauge couplings at $\MI$, unless one departs from the ``survival hypothesis'' and allow intermediate-mass fields. We therefore consider the flipped $\SU(5)$ model, \textit{i.e.} $\G{51}=\SU(5)\times\text{U}(1)_X$~\cite{DeRujula:1980qc,Barr:1981qv,Derendinger:1983aj,Antoniadis:1987dx,Ellis:1988tx}, in which the mixing between the external $\text{U}(1)_X$ and the Abelian charge from inside $\SU(5)$ to produce the hypercharge has the potential to help achieve gauge coupling unification.

The model which we construct, motivated by minimality, is one in which the $\SO10$ symmetry is broken by a vev in the $\rep{45}_H$. In order to break the symmetry down to $\G{321}$, one can use the $\rep{24}_0$ from within the $\rep{45}_H$ together with the $\repb{50}_{2}$ from within the $\repb{126}_H$.

Between $\MGUT$ and $\MI$, the fermions are embedded as $\rep{10}_1\oplus \repb{5}_{-3}\oplus\rep{1}_5$. The scalars are $\rep{24}_0$ from the $\rep{45}_H$ for the breaking together with $\repb{50}_{2}$ for the breaking and $\rep{45}_{-2}$ for the $\SU(2)_\text{L}$ doublet from within the $\repb{126}_H$. From the $\rep{10}_H$, we have $\rep{5}_{-2}$ also for the SM Higgs. The resulting $\beta$ coefficients can be found in App.~\ref{app:beta}.

To compute the RG running, the Abelian charge must be normalized by a factor of $1/\sqrt{40}$. Normalizing also the hypercharge by its usual GUT factor, the matching conditions for the gauge couplings at $\MI$ can be derived. One must also take into account that since the $\SU(5)$ group contains the $\SU(3)_\text{C}\times\SU(2)_\text{L}$ part of $\G{321}$, these must match at $\MI$. The unification of $\alpha_2^{-1}$ and $\alpha_3^{-1}$ therefore determines $\MI$. Hence, the matching conditions read
\begin{align}
\alpha_5^{-1}(\MI) &= \alpha_3^{-1}(\MI) = \alpha_2^{-1}(\MI), \\
\alpha_{1X}^{-1}(\MI) &= \frac{25}{24}\alpha_Y^{-1}(\MI) - \frac{1}{24} \alpha_2^{-1}(\MI).
\end{align}

Using these matching conditions, the RG running may be computed and is displayed in Fig.~\ref{fig:51}. As is shown, gauge coupling unification is not achieved in this model due to the diverging lines of $\alpha_5^{-1}$ and $\alpha_{1X}^{-1}$. To rectify this, the model would need to be made more complicated in order to either significantly change the RG running between $\MI$ and $\MGUT$ or to change the RG running in the SM region so as to change $\MI$.

\subsection{Gauge Coupling Running}
The RG running of the gauge couplings for all models discussed in Sec.~\ref{sec:models} are displayed in Fig.~\ref{fig:RGrunning}. In this figure, the inverse gauge couplings $\alpha_i^{-1}$ are plotted as functions of the energy scale $\mu$. Results to two-loop (one-loop) order are shown by the solid (dashed) lines and the corresponding scales that result in gauge coupling unification are displayed as vertical lines. For the two models in which gauge coupling unification is not achieved, namely the models based on $\G{3211}$ and $\G{51}$, representative intermediate scales are chosen. Particularly for the model based on $\G{51}$, this corresponds to the scale at which the gauge couplings corresponding to $\SU(3)_\text{C}$ and $\SU(2)_\text{L}$ unify. The gauge coupling that each color corresponds to is given by the label in the figure. The subscript of each $\alpha^{-1}$ denotes which of the gauge group factors it corresponds to.

\begin{figure}[t]
\centering
\subfloat[\label{fig:SM}$\G{321}$]{\includegraphics[width=.48\linewidth]{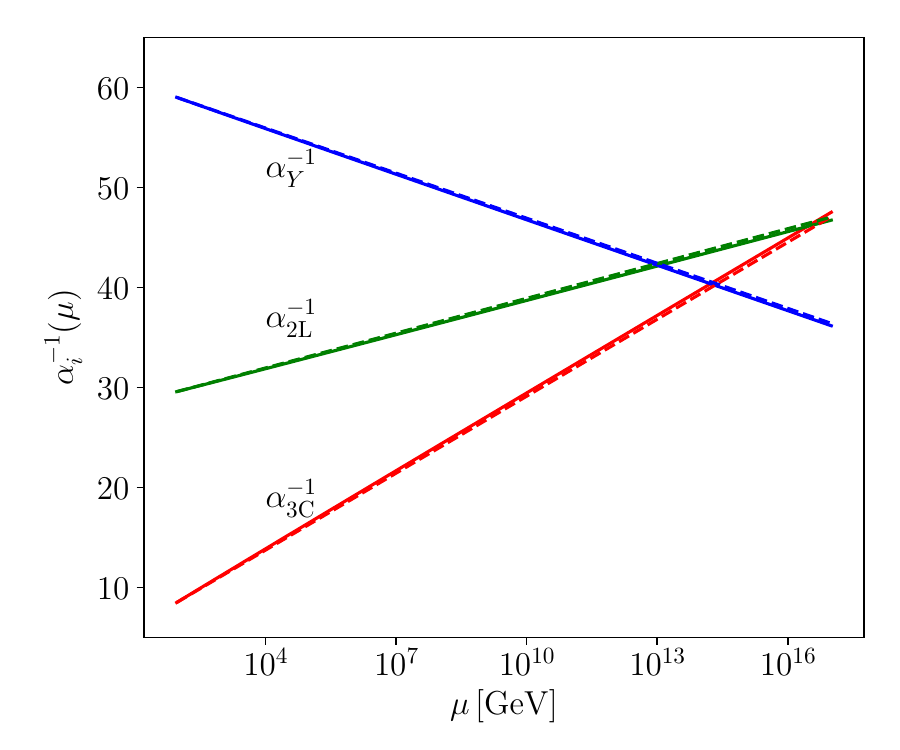}}\,
\subfloat[\label{fig:PS}$\G{422}$]{\includegraphics[width=.48\linewidth]{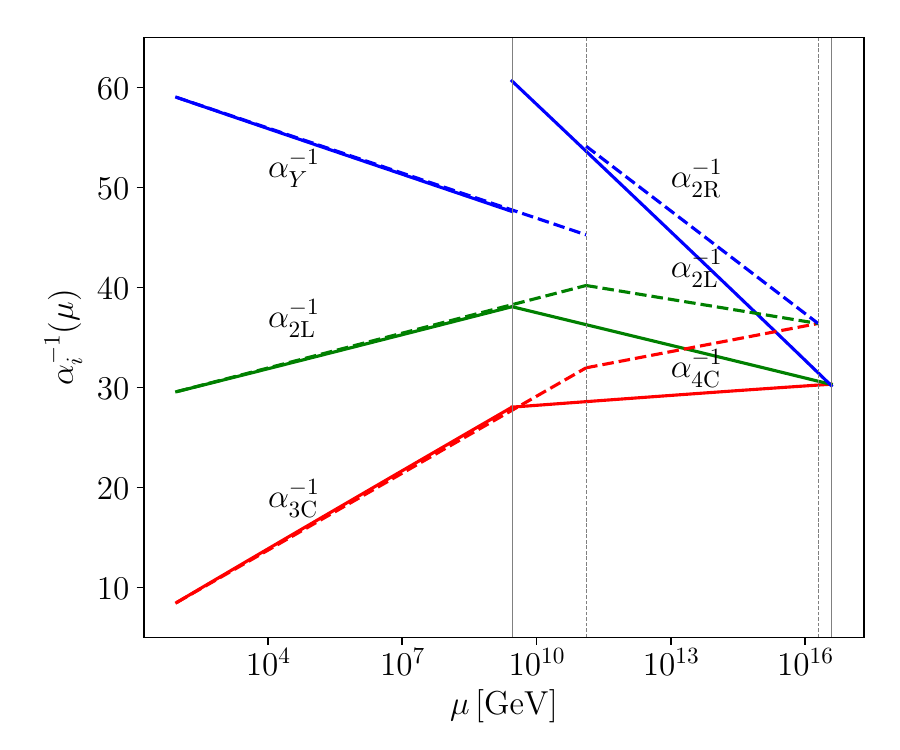}}\,
\subfloat[\label{fig:PSD}$\G{422D}$]{\includegraphics[width=.48\linewidth]{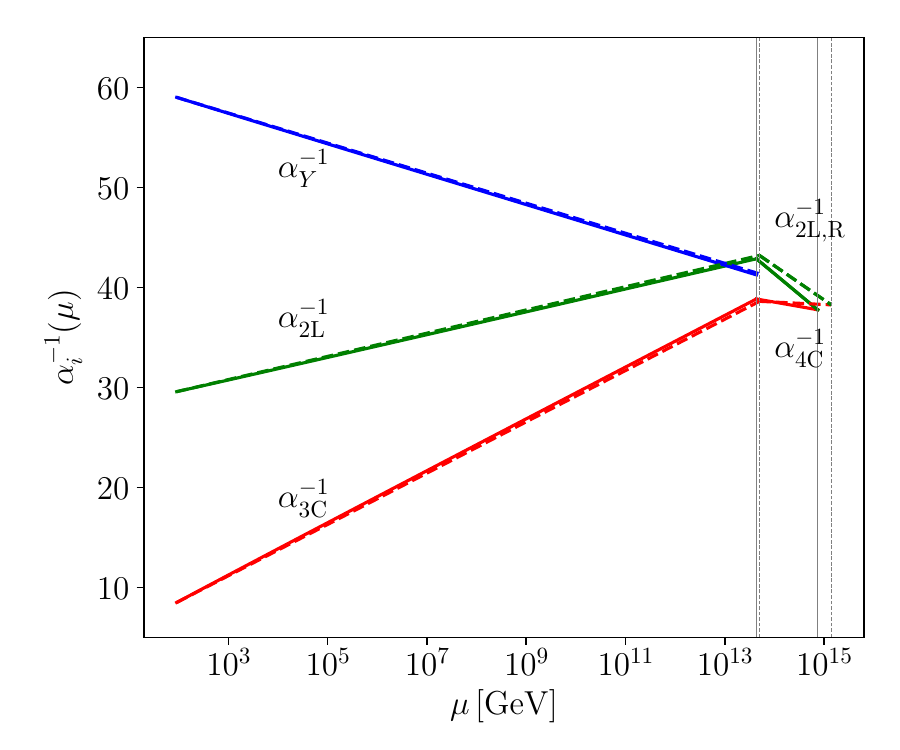}}\,
\subfloat[\label{fig:421}$\G{421}$]{\includegraphics[width=.48\linewidth]{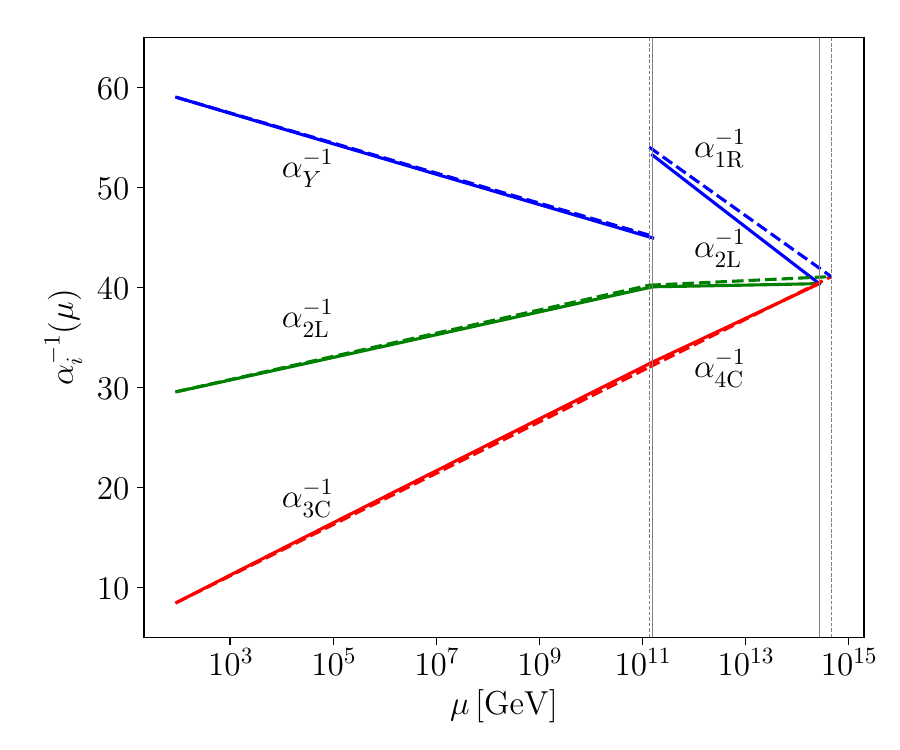}}\,

\caption{\label{fig:RGrunning}RG running of the inverse gauge couplings in the eight different models considered. The results to two-loop (one-loop) order are shown as solid (dashed) lines. For the models without gauge coupling unification, representative values for the scales were chosen. GUT normalization of hypercharge is used throughout. (Continued on next page)}
\end{figure}

\begin{figure}\ContinuedFloat
\centering
\subfloat[\label{fig:3221}$\G{3221}$]{\includegraphics[width=.48\linewidth]{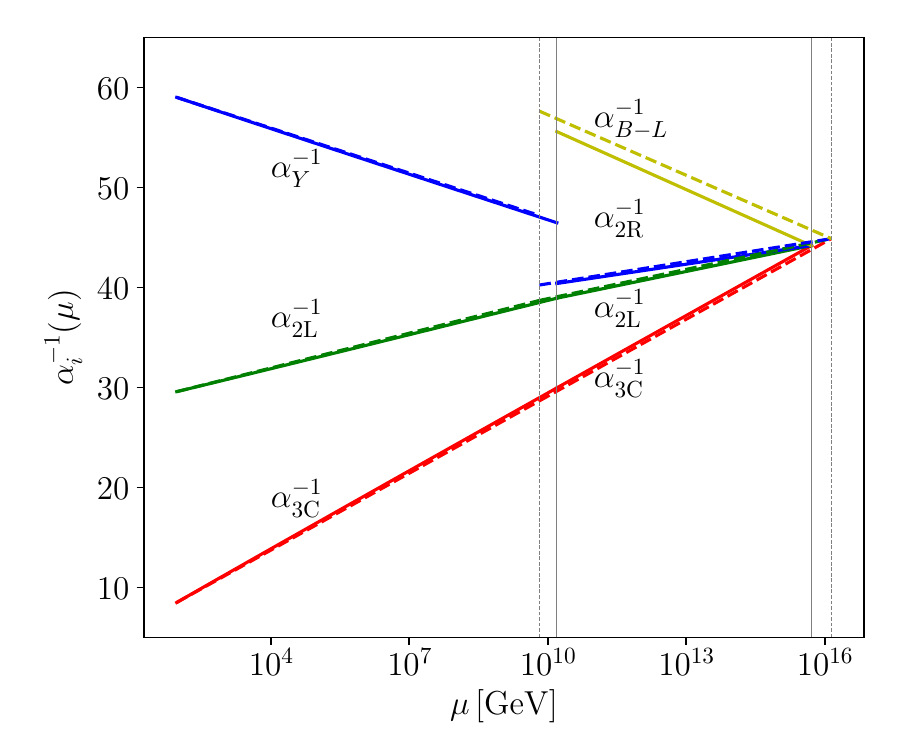}}\,
\subfloat[\label{fig:3221D}$\G{3221D}$]{\includegraphics[width=.48\linewidth]{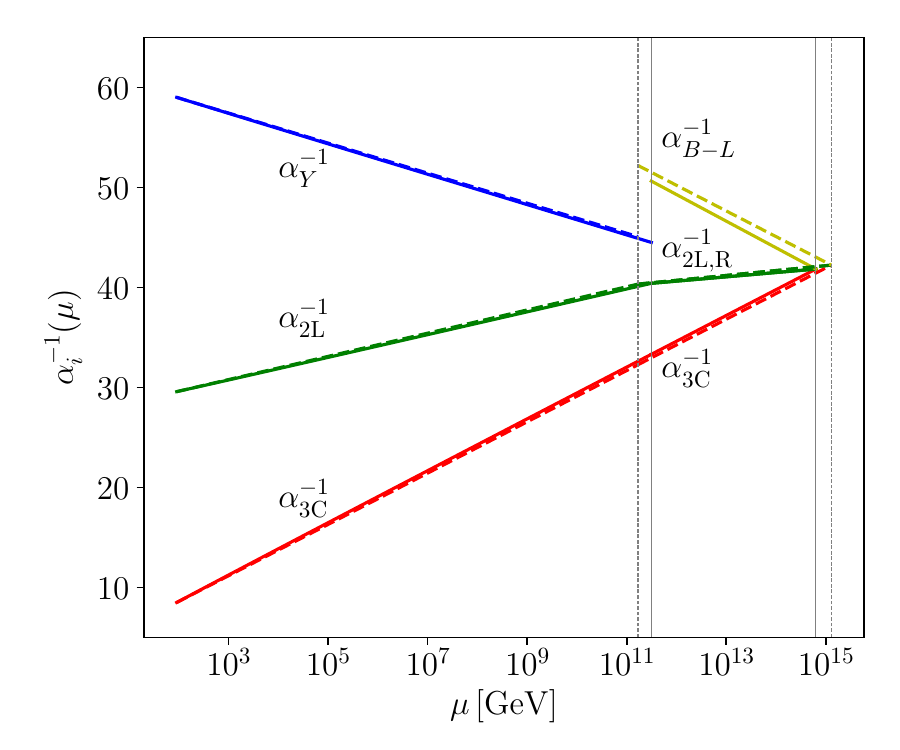}}\,
\subfloat[\label{fig:3211}$\G{3211}$]{\includegraphics[width=.48\linewidth]{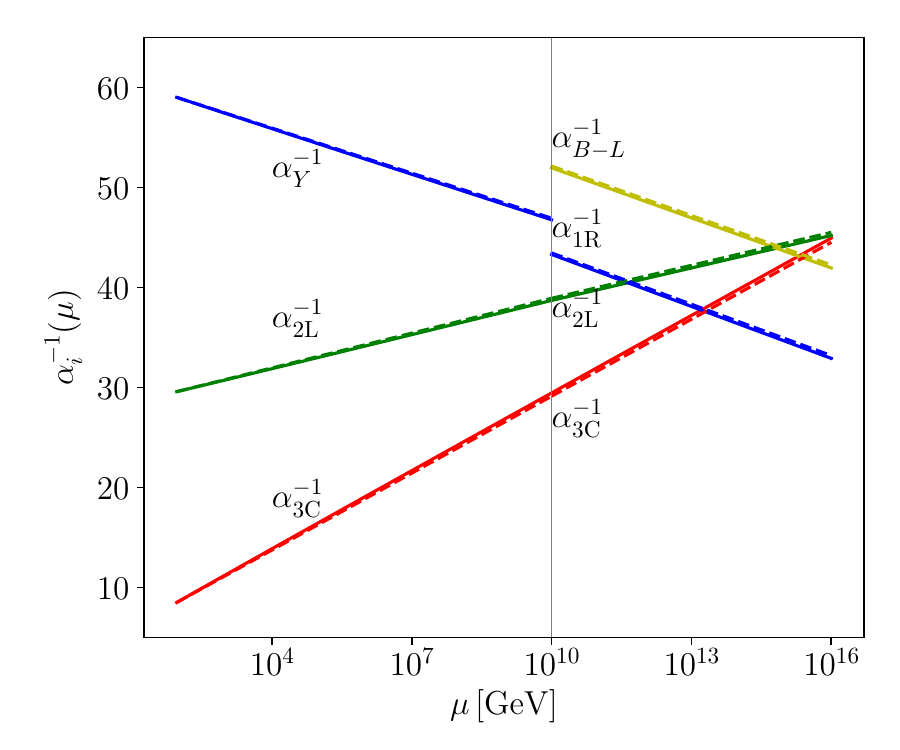}}\,
\subfloat[\label{fig:51}$\G{51}$]{\includegraphics[width=.48\linewidth]{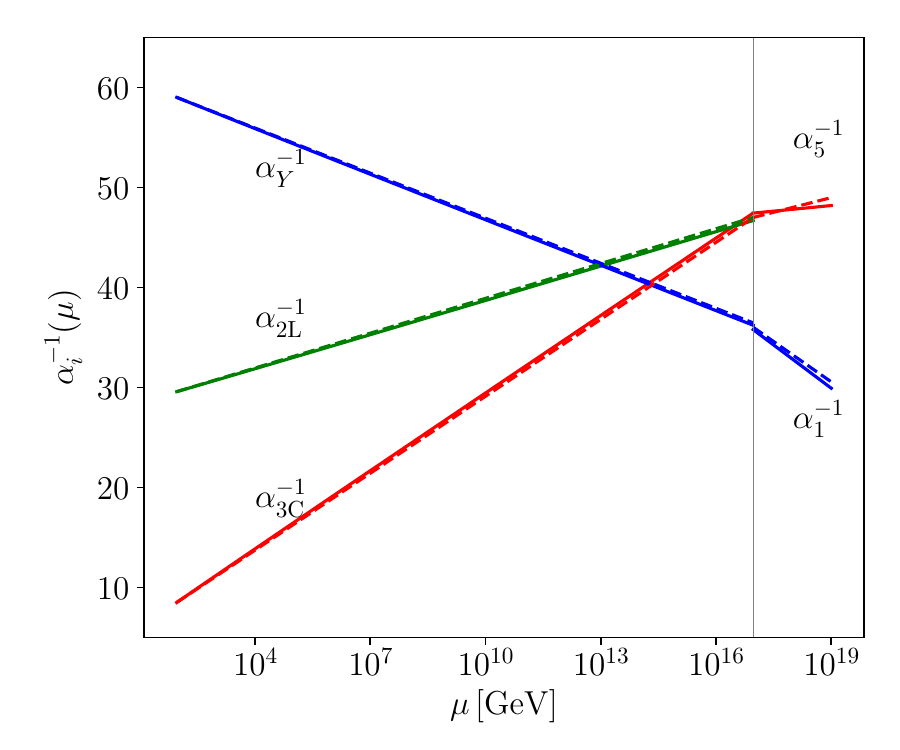}}

\caption{(Continued from previous page) RG running of the inverse gauge couplings in the eight different models considered. The results to two-loop (one-loop) order are shown as solid (dashed) lines. For the models without gauge coupling unification, representative values for the scales were chosen. GUT normalization of hypercharge is used throughout.}
\end{figure}

\section{Threshold Corrections}\label{sec:thresholds}
The results presented in Sec.~\ref{sec:models} assume that the matching of two models occurs at tree-level, meaning that the gauge couplings of the subgroup are equal to a linear combination of the gauge couplings of the group from which it originates. At higher-loop orders, the matching conditions are modified by threshold corrections.

For the symmetry breaking of a group $\mathcal{G}_m$ to another group $\mathcal{G}_n$ at a scale $M_{m\rightarrow n}$, the matching condition with threshold corrections reads
\begin{equation}\label{eq:matching}
\alpha_n^{-1}(M_{m\rightarrow n}) = \alpha_m^{-1}(M_{m\rightarrow n}) - \frac{\lambda^m_n}{12\pi},
\end{equation}
where the one-loop threshold corrections $\lambda^m_n$ are given by~\cite{Hall:1980kf,Weinberg:1980wa}
\begin{equation}\label{eq:thresholds}
\lambda^m_n = \sum_{i\, \in\, \text{vectors}}k_{V_i} S_2(V_i) + \sum_{i\, \in\, \text{scalars}} \kappa_{S_i} k_{S_i} S_2(S_i) \ln\left( \frac{M_{S_i}}{M_{m\rightarrow n}}\right).
\end{equation}
Here, the $\kappa_{S_i}$ are $1$ or $2$ for real or complex representations, while $k_{V_i}$ and $k_{S_i}$ are the multiplicities of the vector and scalar field, respectively, taking into account the dimension of the representation under the other gauge group factors. Note that we assume that all superheavy vector bosons have masses which coincide with the symmetry breaking scale so that there is no scale-dependent term for vectors. If there were superheavy fermions, they would also contribute to the threshold corrections. In what follows, we use the shorthand notation $\eta_i = \ln (M_{S_i}/M_{m\rightarrow n})$.

Note that Eq.~\eqref{eq:matching} holds when the matching of the gauge couplings at tree-level is such that they are equal at the scale of symmetry breaking. This is not the case, for example, when breaking the $\G{422}$ symmetry to $\G{321}$, in which case the hypercharge generator is a linear combination of one generator from $\SU(4)_\text{C}$ and one from $\SU(2)_\text{R}$. The matching of the gauge couplings in that and similar models involves forming a linear combination of the gauge couplings of the broken symmetry group. To each such term, one adds the threshold correction. For more explicit details, we refer the reader to Ref.~\cite{Bertolini:2009qj}.

Given the fields that lie around each energy scale as given in App.~\ref{app:fields}, the threshold corrections to each of the gauge couplings may be computed using Eq.~\eqref{eq:thresholds}. Since there is one threshold correction corresponding to each gauge group factor in the unbroken symmetry at the symmetry breaking scale, one needs to use the representations under the unbroken group of all the fields that lie at the symmetry breaking scale. For each of the models in Sec.~\ref{sec:models} that achieve gauge coupling unification and the model without an intermediate symmetry group, the threshold corrections to the matching conditions have been computed and are presented in App.~\ref{app:thresholds}. Note that, in this work, we do not analyze the effects of threshold corrections for the models based on $\G{3211}$ and $\G{51}$ since these do not achieve gauge coupling unification.

\section{Results}\label{sec:results}
Given the threshold corrections given in App.~\ref{app:thresholds}, we randomly sample the masses of the scalars around the symmetry breaking scale and thereby find how large the deviations from the symmetry breaking scale are required to be in order to save the models. 

\begin{figure}[t]
\centering
\includegraphics[width=0.8\textwidth]{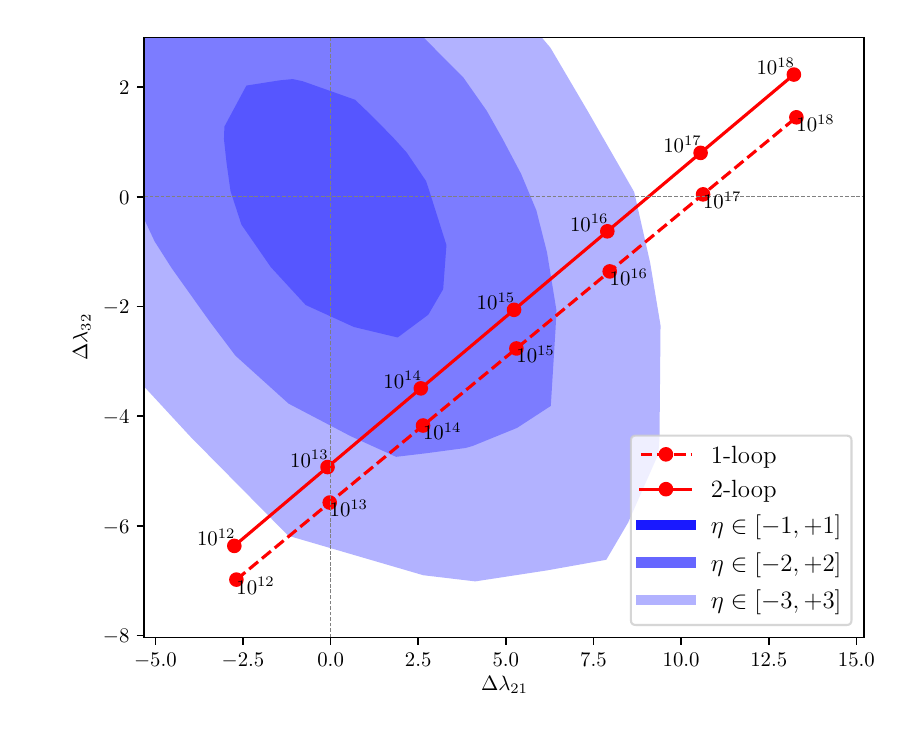}
\caption{\label{fig:SMfailure} Threshold effects to allow for unification of the SM gauge couplings. The difference between gauge couplings $\alpha_3^{-1}-\alpha_2^{-1}$ and $\alpha_2^{-1}-\alpha_1^{-1}$ is shown in the solid (dashed) red line at two-loop (one-loop) level. Numbers above the lines are the energy scales in units of $\text{GeV}$. The blue shaded regions show the size of the threshold corrections. Intersections between the blue shaded regions and the red lines correspond to successful gauge coupling unification.}
\end{figure}

In the case of no intermediate symmetry, the impact of the threshold effects on the matching conditions is such that they can compensate for the difference between the gauge couplings and therefore allow gauge coupling unification~\cite{Ellis:2015jwa,Schwichtenberg:2018cka}. That is, since threshold corrections are meant to account for the failure of gauge coupling unification, we can compare the difference of the gauge couplings with the size of the threshold corrections. To this end, we define
\begin{equation}
\Delta\lambda_{ij}(\mu) = \alpha^{-1}_i(\mu) - \alpha^{-1}_j(\mu) = \frac{\lambda_j^{10} - \lambda_i^{10}}{12\pi}.
\end{equation}
From the three gauge couplings in the SM, the failure of gauge coupling unification can be demonstrated by the two differences $\Delta\lambda_{32}$ and $\Delta\lambda_{21}$. For each energy scale, we can plot the correlation of these two quantities, as is shown by the red lines in Fig.~\ref{fig:SMfailure}, in which the solid line shows the result with RG running at two-loop level and the dashed line at one-loop level. These lines demonstrate the size of the threshold corrections that would be required in order to obtain gauge coupling unification. 

From the expressions for the threshold corrections given in App.~\ref{app:thresholds_SM}, one may compute the size of the difference of the threshold corrections, given values of the parameters $\eta_i$. We randomly sample these parameters independently according to uniform distributions in the regions $\eta_i\in[-1,+1]$, $\eta_i\in[-2,+2]$, and $\eta_i\in[-3,+3]$. These regions demonstrate the possible size of threshold corrections that can be obtained by allowing each scalar mass to vary within the determined ranges.  Although the masses of the scalar fields are in general related, the scalar potential of realistic $\SO10$ models is often complicated and involves many free parameters. Therefore, there is a significant amount of freedom and we assume that the masses are independent.

\begin{figure}[!htbp]
\centering
\subfloat[\label{fig:thr_422}$\G{422}$]{\includegraphics[width=.48\linewidth]{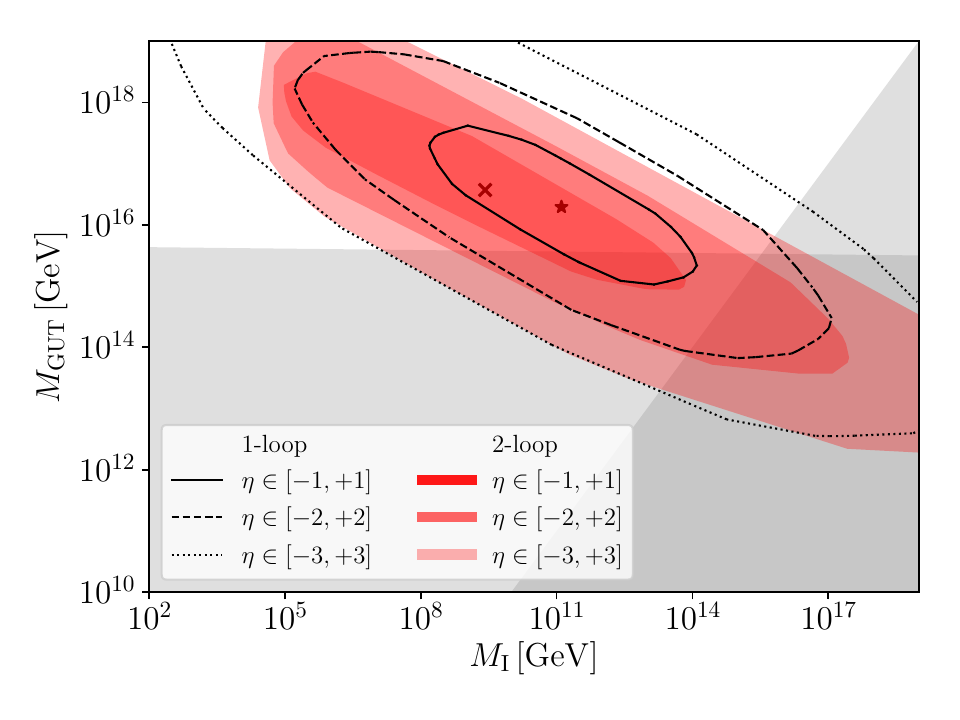}}
\subfloat[\label{fig:thr_422D}$\G{422D}$]{\includegraphics[width=.48\linewidth]{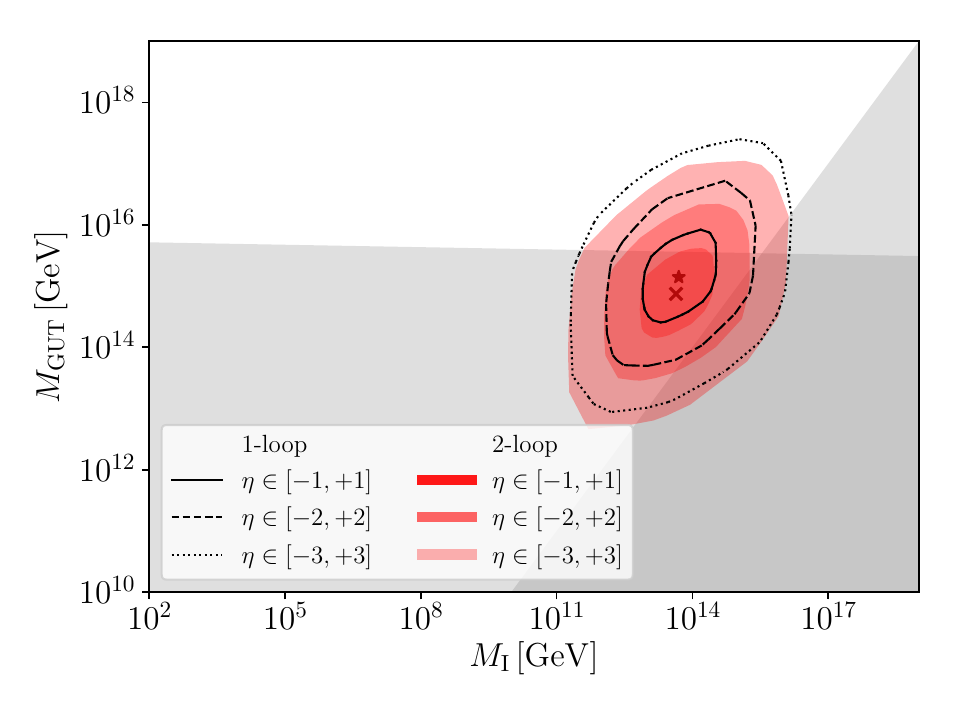}}\\[-1ex]
\subfloat[\label{fig:thr_421}$\G{421}$]{\includegraphics[width=.48\linewidth]{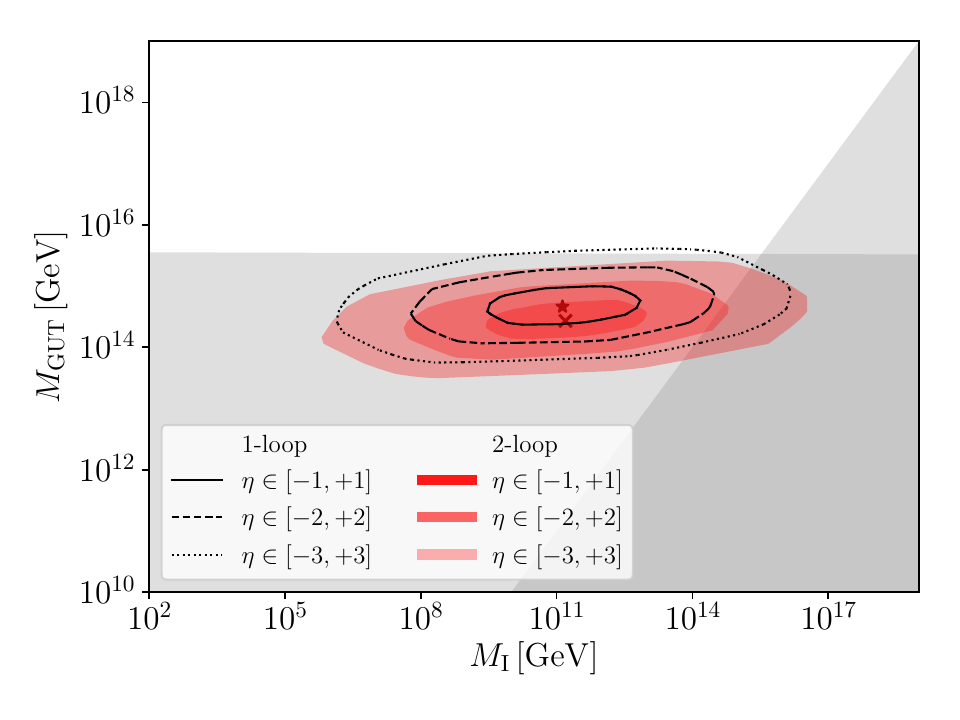}}\\[-1ex]
\subfloat[\label{fig:thr_3221}$\G{3221}$]{\includegraphics[width=.48\linewidth]{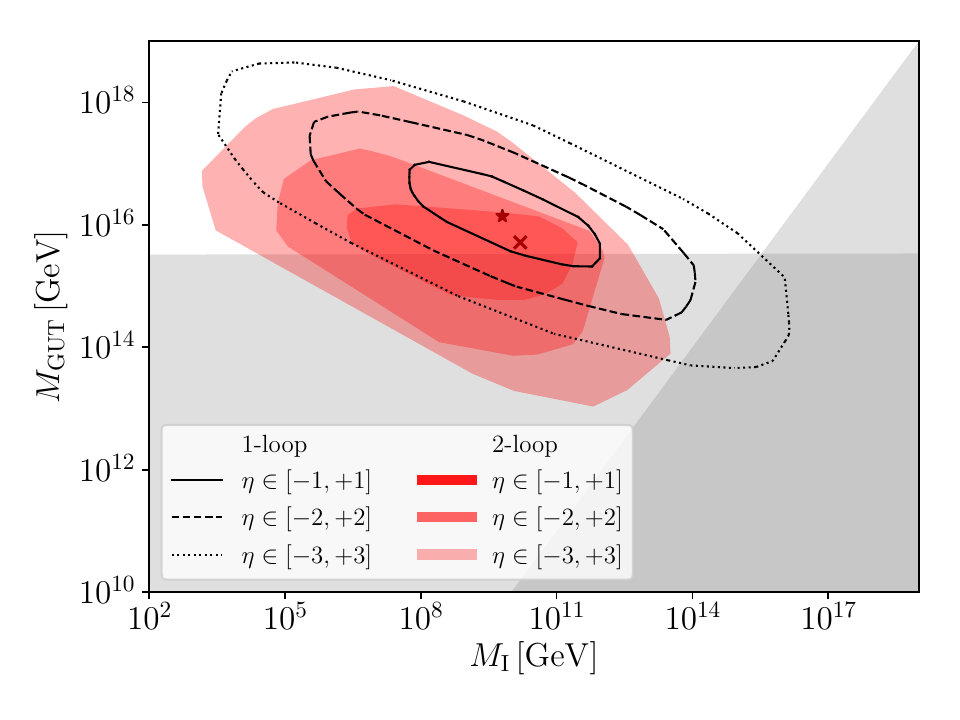}}
\subfloat[\label{fig:thr_3221D}$\G{3221D}$]{\includegraphics[width=.48\linewidth]{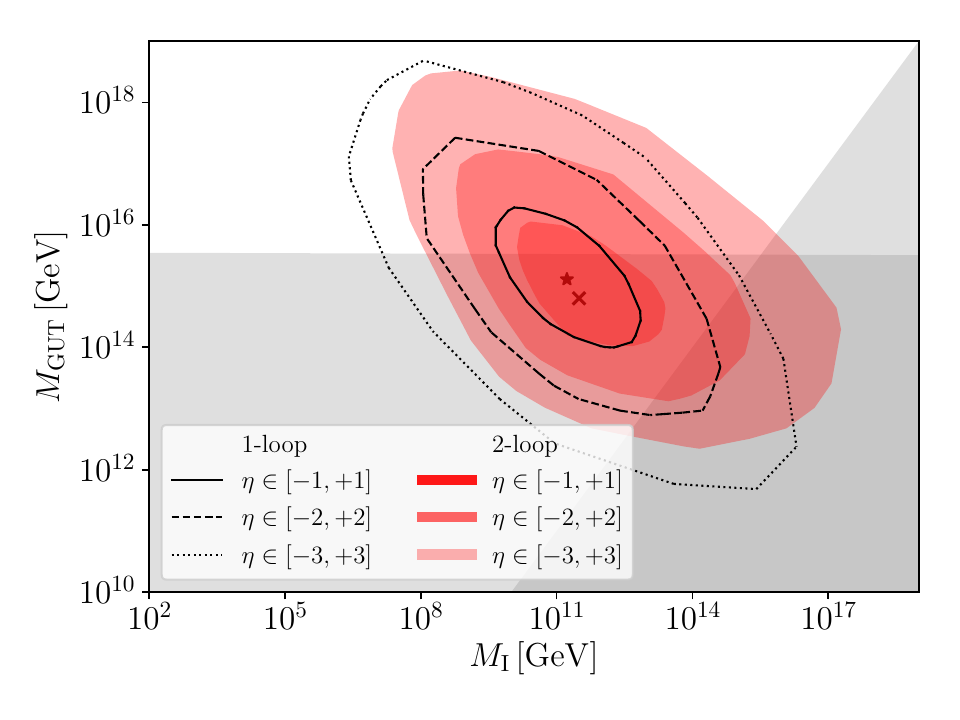}}

\caption{\label{fig:thresholds}Variations in the scales due to threshold corrections for the models which achieve gauge coupling unification. The red shaded regions show the results with RG running at two-loop level, while the full, dashed, and dotted contours show the results at one-loop level. Concentric regions show the scales for the thresholds in the range $\eta_i\in [- 1,+1], [-2,+2]$, and $[-3,+3]$, respectively. The scales without threshold corrections are marked with a ``$\times$" (``$\star$") for the two-loop (one-loop) result. The grey shaded regions are forbidden by a too short proton lifetime and $\MI > \MGUT$.}
\end{figure}

Computationally, each scan was performed on a computing cluster utilizing 48 cores each sampling at least $2\times 10^6$ points.\footnote{The actual number of points varied between the models due to different numerical complexity involved in solving the system of equations. The two-loop results of $\G{3221}$ was generated by sampling $2\times 10^6$ points per core, the two-loop results of $\G{422}$ and $\G{3221D}$ by sampling $1\times 10^7$ points per core, while all other results were generated by sampling $8\times 10^8$ points per core.} Following this, the convex hull was computed using the \verb|ConvexHull| routine from the \verb|SciPy|~\cite{2020SciPy-NMeth} package version \verb|1.3.2| in \verb|Python 3.7.0|, which produced the blue shaded regions shown in Fig.~\ref{fig:SMfailure}. This was performed by first sampling $10^4$ points for each region and using the \verb|ConvexHull| routine to find the smallest convex polygon containing the given points. From there, only the points on the boundary were saved. When sampling the points, only those that fell outside the initial boundary were saved and were used to create the new boundary. This produced the regions shown in Figs.~\ref{fig:SMfailure} and~\ref{fig:thresholds}. These figures were used to illustrate the results rather than scatter plots, since they better demonstrate the fact that the sampled threshold corrections span a region.

The intersection of the blue shaded regions and the red lines shows the size of the threshold corrections that lead to unification of the SM gauge couplings at a particular scale. For example, with RG running at two-loop level, threshold corrections with $\eta_i\in[-2,+2]$ can provide unification, but only at a range of scales around $\MGUT\simeq (5\times 10^{13} - 5\times 10^{15})\GeV$, with a slightly narrower range at one-loop level. This corresponds to a proton lifetime of $\tau_p\approx 5\times 10^{34}\,\text{yr}$, which is allowed but close to the current bound. With $\eta_i\in[-3,+3]$, a larger range of values of $\MGUT$ are allowed, and gauge coupling unification can then be successfully achieved.

For the models with an intermediate symmetry that achieve gauge coupling unification without threshold corrections, the effect of threshold corrections on the scales $\MI$ and $\MGUT$ were found. This was performed by solving the matching conditions to determine the two scales as functions of the $\eta_i$ parameters. Then, the $\eta_i$ were individually sampled in the same way as described above for Fig.~\ref{fig:SMfailure}. Regions of possible scales for $\eta_i\in[-1,+1]$, $\eta_i\in[-2,+2]$, and $\eta_i\in[-3,+3]$ are plotted in Fig.~\ref{fig:thresholds}, again using the \verb|ConvexHull| routine. The red shaded regions denote the range of possible scales with RG running at two-loop level and the black contours show the same at one-loop level. The scales obtained in the absence of threshold corrections are denoted by a ``$\times$" (``$\star$") for the two-loop (one-loop) result.

To investigate if the models which were originally disfavored due to a too short proton lifetime can be saved by threshold corrections, the nearly horizontal grey shaded region denotes a too short proton lifetime. Thus, points in the shaded region are ruled out. This line was computed by first calculating the proton lifetime for the randomly sampled points and then using a Support Vector Machine (SVM) from the \verb|scikit-learn| package~\cite{scikit-learn} version \verb|0.19.2| in \verb|Python 3.7.0| to find the equation of the line that best separates the two classes. This line separates the two classes well and was observed to be nearly identical for the one-loop and two-loop RG running. The slanted shaded region is the forbidden region corresponding to $\MI > \MGUT$. Note that there also exist bounds on $\MI$, such as right-handed neutrino masses in the type I seesaw mechanism or leptogenesis~\cite{Fukugita:1986hr,Covi:1996wh,Davidson:2002qv,Buchmuller:2004nz,Blanchet:2008pw,Fong:2014gea}. Other possible experimental constraints on GUTs may come from neutron-antineutron oscillations~\cite{Mohapatra:1980qe,Mohapatra:1989ze,Mohapatra:2009wp,Babu:2013jba,Phillips:2014fgb} induced by the breaking of $B-L$ or topological defects~\cite{Lazarides:1980cc,Lazarides:1980va,Lazarides:1981fv,Kibble:1982ae,Kibble:1982dd,Weinberg:1983bf,Bhattacharjee:1991zm,Davis:1994py,Chakrabortty:2017mgi,Chakrabortty:2019fov,Lazarides:2019xai} from the breaking of some of the symmetries considered. The stability of these topological defects varies between different models, making some more problematic than others. For more details, see \textit{e.g.} Refs.~\cite{Lazarides:1980cc,Lazarides:1981fv,Chakrabortty:2017mgi,Chakrabortty:2019fov}

For the model based on $\G{422}$, it was already allowed without threshold corrections. As shown in Fig.~\ref{fig:thr_422}, they have quite a large effect on the scales. The model based on $\G{422D}$ has smaller threshold corrections, as displayed in Fig.~\ref{fig:thr_422D}, due to the absence of the $\rep{210}_H$. This model was disfavored without threshold corrections and is saved by threshold corrections with $\eta_i\in[-1,+1]$. Turning to the model based on $\G{421}$ shown in Fig.~\ref{fig:thr_421}, it was disfavored without threshold corrections and is still disfavored by a too short proton lifetime with $\eta_i\in[-3,+3]$. It requires $\eta_i\in[-4,+4]$ in order to predict a long enough proton lifetime. The model based on $\G{3221}$ shown in Fig.~\ref{fig:thr_3221} was also allowed without threshold corrections. Lastly, the model based on $\G{3221D}$, shown in Fig.~\ref{fig:thr_3221D}, predicted a too short proton lifetime in the absence of threshold corrections. Only a small amount of threshold corrections are required to save it, since part of the region with $\eta_i\in[-1,+1]$ is above the grey shaded region. We summarize these findings in Tab.~\ref{tab:summary}. 

\begin{table}[t]
\caption{\label{tab:summary}Comparison of the viability of the models considered. The second column (``GCU") denotes whether or not gauge coupling unification is achieved without threshold corrections. The next four columns show if the predicted proton lifetime is above the lower bound for various sizes of threshold corrections. A checkmark (``\cmark ") denotes an allowed model whereas a cross (``\xmark ") denotes a disfavored model. The dashes in the last two rows signify that we did not investigate threshold corrections in those models. The results given are with RG running at two-loop level. Results at one-loop level may be found in the main text and the figures.}
\centering
\begin{tabular}{l c c c c c}
\hline
\hline
Model & GCU & $\eta_i\in[-3,+3]$ & $\eta_i\in[-2,+2]$ & $\eta_i\in[-1,+1]$ & no thresh.\\
\hline
$\G{321}$ & \xmark & \cmark & \cmark & \xmark & \xmark\\
$\G{422}$ & \cmark & \cmark & \cmark & \cmark & \cmark \\
$\G{422D}$ & \cmark & \cmark & \cmark & \cmark & \xmark \\
$\G{421}$ & \cmark & \xmark & \xmark & \xmark & \xmark\\
$\G{3221}$ & \cmark & \cmark & \cmark & \cmark & \cmark \\
$\G{3221D}$ & \cmark & \cmark & \cmark & \cmark & \xmark \\
$\G{3211}$ & \xmark & {\bf --} & {\bf --} & {\bf --} & {\bf --}\\
$\G{51}$ & \xmark & {\bf --} & {\bf --} & {\bf --} & {\bf --}\\
\hline
\hline
\end{tabular}
\end{table}

\section{Summary and Conclusions}\label{sec:conclusion}
Unification in models in which the $\SO10$ gauge symmetry breaks to the SM directly or with one intermediate symmetry breaking have been investigated. Particularly, we have considered non-SUSY models, in which it is known that achieving gauge coupling unification is more difficult than in their SUSY counterparts. 

We have solved the RG running in these models and studied whether or not gauge coupling unification is achieved and, if so, whether or not the prediction for the proton lifetime is above the experimental lower bound. Gauge coupling unification was achieved in models with $\G{422}$, $\G{422D}$, $\G{421}$, $\G{3221}$, or $\G{3221D}$ as intermediate symmetries. Of these, only $\G{422}$ and $\G{3221}$ predicted a proton lifetime that is above the experimental lower bound, with $\tau_p\approx 1.2\times 10^{38}\,\text{yr}$ and $\tau_p \approx 9.4\times 10^{34}\,\text{yr}$, respectively. These two models are the well-studied PS and left-right models.

Threshold corrections have been computed for the model with direct breaking as well as the models in which gauge coupling unification occurred. For the model with direct breaking, we have found that gauge coupling unification can be achieved with a long enough proton lifetime if $\eta_i\in[-2,+2]$. This holds with RG running at both one-loop and two-loop level.

The models with intermediate symmetries for which the predicted proton lifetimes are too short can be saved by invoking threshold corrections, given that they are large enough. For the models based on $\G{422D}$ and $\G{3221D}$, it is sufficient to have $\eta_i\in[-1,+1]$, both with one-loop and two-loop level RG running. The model based on $\G{421}$, on the other hand, requires larger threshold corrections. With RG running at one-loop level, the proton decay bound can be evaded with $\eta_i\in[-3,+3]$, while at two-loop level, $\eta_i\in [-4,+4]$ is required.  It should be noted that already for $\eta_i\in [-3,+3]$, the perturbation from the relevant scale is quite large (a factor of about $20$). Such a large deviation may therefore not be considered to be very natural, since a significant amount of fine-tuning may be necessary. The main point of this work is to illustrate this trade-off between naturalness and viability in the various models based on $\SO10$.

The results in this work assume the specific model details as described in Sec.~\ref{sec:models}. It should be noted that it is possible to modify some of these details while still achieving the same symmetry breaking chain. However, the results reported in this work can be seen as representative of these models. Furthermore, in the construction of the models, we have not taken into account any constraints from the scalar potential. It would be interesting to investigate this since there can be correlations between the masses of the scalars that could impact the results.

Additionally, we have not taken into account any bounds on physics related to $\MI$. This may in some of the models be related to neutrino masses and/or leptogenesis. Furthermore, we have neglected the effect that perturbing the scalar fields around $\MI$ has on the RG running, as investigated for example in Ref.~\cite{Babu:2015bna}. For larger perturbations, this effect may become substantial and have an effect on the conclusions. Another effect that may have an impact on our results is the existence of Planck-suppressed higher-dimensional operators, especially when $\MGUT$ is large. As discussed in Refs.~\cite{Shafi:1983gz,Hill:1983xh,Calmet:2008df}, such operators can modify the field strength tensor of the unbroken subgroup and hence have an effect on the matching conditions of the gauge couplings. Furthermore, the inclusion of Planck-suppressed higher-dimensional operators can affect the proton decay rate~\cite{Harnik:2004yp,Barr:2012xb}. The above mentioned points, together with the investigation of the two models that were not considered here, namely those based on $\G{3211}$ and $\G{51}$, would make for an interesting future study.

\appendix

\section{Beta Coefficients}
\label{app:beta}
The $\beta$ coefficients for the models discussed in Sec.~\ref{sec:models} are listed in Tab.~\ref{tab:beta}. The second column lists the $\beta$ functions at one-loop level and the third column lists them at two-loop level.

\needspace{3\baselineskip}
\begin{longtable}{l c c}
\caption{\label{tab:beta} $\beta$ coefficients at one-loop and two-loop levels in each of the eight models considered. The order in which the $\beta$ functions for each model are listed is the same order in which the gauge group factors are listed, following the conventions in the rest of this work.}
\endhead
\hline\hline
Model & 1-loop level & 2-loop level \\
\hline
$\G{321}$ & $(-7, -\frac{19}{6}, \frac{41}{10})$ & $\begin{pmatrix}
-26 & \frac{9}{2} & \frac{11}{10} \\
12 & \frac{35}{6} & \frac{9}{10} \\
\frac{44}{5} & \frac{27}{10} & \frac{199}{50}
\end{pmatrix}$\\
$\G{422}$ & $(-\frac{7}{3}, 2, \frac{28}{3})$ & $\begin{pmatrix}
\frac{2435}{6} & \frac{105}{2} & \frac{249}{2} \\
\frac{525}{2} & 73 & 48 \\
\frac{1245}{2} & 48 & \frac{835}{3}
\end{pmatrix}$\\
$\G{422D}$ & $(\frac{2}{3}, \frac{28}{3}, \frac{28}{3})$ & $\begin{pmatrix}
\frac{3551}{6} & \frac{249}{2} & \frac{249}{2} \\
\frac{1245}{2} & \frac{835}{3} & 48 \\
\frac{1245}{2} & 48 & \frac{835}{3}
\end{pmatrix}$\\
$\G{421}$ & $(-7, -\frac{2}{3}, 10$) & $\begin{pmatrix}
\frac{265}{2} & \frac{57}{2} & \frac{43}{2} \\
\frac{285}{2} & \frac{115}{3} & 8 \\
\frac{645}{2} & 24 & 51
\end{pmatrix}$\\
$\G{3221}$ & $(-7, -\frac{8}{3}, -2, \frac{11}{2})$ & $\begin{pmatrix}
-26 & \frac{9}{2} & \frac{9}{2} & \frac{1}{2} \\
12 & \frac{37}{3} & 6 & \frac{3}{2} \\
12 & 6 & 31 & \frac{27}{2} \\
4 & \frac{9}{2} & \frac{81}{2} & \frac{61}{2}
\end{pmatrix}$\\
$\G{3221D}$ & $(-7, -\frac{4}{3}, -\frac{4}{3}, 7)$ & $\begin{pmatrix}
-26 & \frac{9}{2} & \frac{9}{2} & \frac{1}{2} \\
12 & \frac{149}{3} & 6 & \frac{27}{2} \\
12 & 6 & \frac{149}{3} & \frac{27}{2} \\
4 & \frac{81}{2} & \frac{81}{2} & \frac{115}{2}
\end{pmatrix}$\\
$\G{3211}$ & $(-7, -3, \frac{14}{3}, \frac{9}{2})$ & $\begin{pmatrix}
-26 & \frac{9}{2} & \frac{3}{2} & \frac{1}{2} \\
12 & 8 & 1 & \frac{3}{2} \\
12 & 3 & 8 & \frac{15}{2} \\
4 & \frac{9}{2} & \frac{15}{2} & \frac{25}{2}
\end{pmatrix}$\\
$\G{51}$ & $(-\frac{8}{3}, \frac{22}{3})$ & $\begin{pmatrix}
\frac{14594}{15} & \frac{129}{10} \\
\frac{1548}{5} & \frac{79}{10}
\end{pmatrix}$\\
\hline\hline
\end{longtable}

\section{Fields at Each Scale}\label{app:fields}
In this appendix, we list the scalar and vector fields that lie at each of the scales in the models considered. Except for the model based on $\G{321}$ which has only one scale, the fields that lie at $\MGUT$ are listed in the third column and the fields that lie at $\MI$ are listed in the forth and fifth columns. The fourth column lists them as representations of the intermediate group and the fifth column lists them as representations of the SM.

The fields for the $\G{321}$ model are listed in Tab.~\ref{tab:SM}, for the $\G{422}$ model in Tab.~\ref{tab:PS}, for the $\G{422D}$ model in Tab.~\ref{tab:PSD}, for the $\G{421}$ model in Tab.~\ref{tab:421}, for the $\G{3221}$ model in Tab.~\ref{tab:3221}, and finally for the $\G{3221D}$ model in Tab.~\ref{tab:3221D}.

\begin{table}[!htb]
\caption{\label{tab:SM} Scalar and vector fields that have masses around $\MGUT$ in the $\G{321}$ model.}
\centering
\begin{tabular}{l l l}
\hline\hline
&$\SO10$ & $\G{321}$  \\
\hline
Scalars & $\rep{10}_H$ & $H_1(\rep{3},\rep{1})_{-1/3}$, $H_2(\repb{3},\rep{1})_{1/3}$, $\phi(\rep{1},\rep{2})_{-1/2}$\\
& $\repb{126}_H$ & $L_1(\rep{1},\rep{3})_{1}, L_2(\repb{3},\rep{3})_{1/3},L_3(\rep{6},\rep{3})_{-1/3}$, $R_1(\rep{3},\rep{1})_{-1/3}, R_2(\rep{3},\rep{1})_{-4/3}, R_3(\rep{3},\rep{1})_{2/3},$\\
& & $R_4(\repb{6},\rep{1})_{4/3}, R_5(\repb{6},\rep{1})_{1/3}, R_6(\repb{6},\rep{1})_{-2/3},$ $R_7(\rep{1},\rep{1})_{-2}, R_8(\rep{1},\rep{1})_{-1}, R_9(\rep{1},\rep{1})_0,$\\
& & $S_1(\rep{3},\rep{1})_{-1/3}, S_2(\repb{3},\rep{1})_{1/3}$, $T_1(\rep{3},\rep{2})_{1/6}, T_2(\rep{3},\rep{2})_{7/6}, T_3(\rep{8},\rep{2})_{-1/2},$ $T_4(\rep{8},\rep{2})_{1/2}$, \\
& & $T_5(\rep{1},\rep{2})_{-1/2}, T_6(\rep{1},\rep{2})_{1/2}, T_7(\repb{3},\rep{2})_{-7/6}, T_8(\repb{3},\rep{2})_{-1/6}$,  \\
& $\rep{144}_H$ & $A_1(\rep{8},\rep{1})_{-1}, A_2(\rep{8},\rep{1})_{0}, A_3(\rep{6},\rep{1})_{-1/3}, A_4(\rep{6},\rep{1})_{2/3}, A_5(\rep{3},\rep{1})_{-1/3}, A_6(\rep{3},\rep{1})_{2/3},$\\
& & $A_7(\repb{3},\rep{1})_{1/3}, A_8(\repb{3},\rep{1})_{4/3}, B_1(\rep{8},\rep{2})_{1/2}, B_2(\repb{6},\rep{2})_{-1/6}, B_3(\rep{3},\rep{2})_{-5/6}, B_4(\repb{3},\rep{2})_{-1/6},$\\
& & $C_1(\rep{3},\rep{3})_{-1/3}, C_2(\rep{3},\rep{3})_{2/3}, C_3(\rep{1},\rep{3})_{-1}, C_4(\rep{1},\rep{3})_{0}, D_1(\repb{3},\rep{2})_{-7/6}, D_2(\repb{3},\rep{2})_{-1/6},$\\
& & $D_3(\repb{3},\rep{2})_{5/6}, D_4(\rep{1},\rep{2})_{-1/2}, D_5(\rep{1},\rep{2})_{1/2}, D_6(\rep{1},\rep{2})_{3/2}, E_1(\rep{3},\rep{1})_{-1/3}, E_2(\rep{3},\rep{1})_{2/3},$ \\
& & $E_3(\rep{1},\rep{1})_{-1}, E_4(\rep{1},\rep{1})_{0}, F_1(\repb{3},\rep{2})_{-1/6}, F_2(\rep{1},\rep{2})_{1/2}$\\
\hline
Vectors & $\rep{45}$ & $(\rep{3},\rep{2})_{-5/6}$, $(\rep{3},\rep{2})_{1/6}$, $(\repb{3},\rep{2})_{-1/6}$, $(\repb{3},\rep{2})_{5/6}$, $(\rep{3},\rep{1})_{2/3}$, $(\repb{3},\rep{1})_{-2/3}$,\\
&&  $(\rep{1},\rep{1})_{-1}$, $(\rep{1},\rep{1})_{0}$, $(\rep{1},\rep{1})_{1}$ \\
\hline\hline
\end{tabular}
\end{table}

\begin{table}[!htb]
\caption{\label{tab:PS}Scalar and vector fields that have masses around $\MGUT$ and $\MI$ in the $\G{422}$ model.}
\centering
\begin{tabular}{l l l l l l}
\hline\hline
&$\SO10$ & $\G{422}$ (fields at $\MGUT$) & $\G{422}$ (fields at $\MI$) & $\G{321}$ (fields at $\MI$) \\
\hline
Scalars &$\rep{10}_H$ & $H(\rep{6},\rep{1},\rep{1})$ & $\Phi(\rep{1},\rep{2},\rep{2})$ & $\phi(\rep{1},\rep{2})_{-1/2}$\\
&$\rep{45}_H$ & $\delta_1(\rep{15},\rep{1},\rep{1}),\delta_2(\rep{1},\rep{3},\rep{1}),$& $\kappa(\rep{1},\rep{1},\rep{3})$ & $\kappa_1(\rep{1},\rep{1})_1, \kappa_2(\rep{1},\rep{1})_{-1}, \kappa_3(\rep{1},\rep{1})_0$\\
&& $\delta_3(\rep{6},\rep{2},\rep{2})$\\
&$\repb{126}_H$ & $\Delta_L(\repb{10},\rep{3},\rep{1}),S(\rep{6},\rep{1},\rep{1})$ & $\Delta_R(\rep{10},\rep{1},\rep{3}),$ & $R_1(\rep{3},\rep{1})_{-1/3}, R_2(\rep{3},\rep{1})_{-4/3}, R_3(\rep{3},\rep{1})_{2/3},$ \\
&& & & $R_4(\repb{6},\rep{1})_{4/3}, R_5(\repb{6},\rep{1})_{1/3}, R_6(\repb{6},\rep{1})_{-2/3},$\\
&&&&$R_7(\rep{1},\rep{1})_{-2}, R_8(\rep{1},\rep{1})_{-1}, R_9(\rep{1},\rep{1})_0$\\
&&& $T(\rep{15},\rep{2},\rep{2})$ & $T_1(\rep{3},\rep{2})_{1/6}, T_2(\rep{3},\rep{2})_{7/6}, T_3(\rep{8},\rep{2})_{-1/2},$\\
&&&& $T_4(\rep{8},\rep{2})_{1/2}, T_5(\rep{1},\rep{2})_{-1/2}, T_6(\rep{1},\rep{2})_{1/2},$\\
&&&& $T_7(\repb{3},\rep{2})_{-7/6}, T_8(\repb{3},\rep{2})_{-1/6}$\\
&$\rep{210}_H$ & $\Sigma_L(\rep{15},\rep{3},\rep{1}),\Sigma_R(\rep{15},\rep{1},\rep{3}),$\\
&&$\xi_1(\rep{10},\rep{2},\rep{2}),\xi_2(\repb{10},\rep{2},\rep{2}),$\\
&&$\xi_3(\rep{15},\rep{1},\rep{1}),\xi_4(\rep{6},\rep{2},\rep{2}),$\\
&&$S^\prime(\rep{1},\rep{1},\rep{1})$\\
\hline
Vectors & $\rep{45}$ & $(\rep{6},\rep{2},\rep{2})$ & $(\rep{15},\rep{1},\rep{1}),$ & $(\rep{3},\rep{1})_{2/3}$, $(\repb{3},\rep{1})_{-2/3}$, $(\rep{1},\rep{1})_0$\\
&&& $(\rep{1},\rep{1},\rep{3})$ & $(\rep{1},\rep{1})_1$, $(\rep{1},\rep{1})_{-1}$ \\
\hline\hline
\end{tabular}
\end{table}

\begin{table}[!htb]
\caption{\label{tab:PSD}Scalar and vector fields that have masses around $\MGUT$ and $\MI$ in the $\G{422D}$ model.}
\centering
\begin{tabular}{l l l l l l}
\hline\hline
& $\SO10$ & $\G{422D}$ (fields at $\MGUT$) & $\G{422D}$ (fields at $\MI$) & $\G{321}$ (fields at $\MI$) \\
\hline
Scalars & $\rep{10}_H$ & $H(\rep{6},\rep{1},\rep{1})$ & $\Phi(\rep{1},\rep{2},\rep{2})$ & $\phi(\rep{1},\rep{2})_{-1/2}$\\
& $\rep{45}_H$ & $\delta_1(\rep{15},\rep{1},\rep{1}),\delta_3(\rep{6},\rep{2},\rep{2})$ & $\kappa(\rep{1},\rep{1},\rep{3}),$ & $\kappa_1(\rep{1},\rep{1})_{-1},\kappa_2(\rep{1},\rep{1})_{0},\kappa_3(\rep{1},\rep{1})_{1}$\\
& & & $\delta_2(\rep{1},\rep{3},\rep{1})$ & $\delta_2(\rep{1},\rep{3})_0$ \\
& $\rep{54}_H$ & $\zeta_0(\rep{1},\rep{1},\rep{1}), \zeta_1(\rep{1},\rep{3},\rep{3}),$ \\
& & $\zeta_2(\rep{6},\rep{2},\rep{2}),\zeta_3(\rep{20^\prime},\rep{1},\rep{1})$\\
& $\repb{126}_H$ & $S(\rep{6},\rep{1},\rep{1})$ & $\Delta_L(\repb{10},\rep{3},\rep{1}),$ & $L_1(\rep{1},\rep{3})_{1}, L_2(\repb{3},\rep{3})_{1/3},L_3(\rep{6},\rep{3})_{-1/3}$ \\
& & & $\Delta_{R}(\rep{10},\rep{1},\rep{3}),$ & $R_1(\rep{3},\rep{1})_{-1/3}, R_2(\rep{3},\rep{1})_{-4/3}, R_3(\rep{3},\rep{1})_{2/3},$\\
& & & & $R_4(\repb{6},\rep{1})_{4/3}, R_5(\repb{6},\rep{1})_{1/3}, R_6(\repb{6},\rep{1})_{-2/3},$\\
& & & &$R_7(\rep{1},\rep{1})_{-2}, R_8(\rep{1},\rep{1})_{-1}, R_9(\rep{1},\rep{1})_0$\\
& & &$T(\rep{15},\rep{2},\rep{2})$ & $T_1(\rep{3},\rep{2})_{1/6}, T_2(\rep{3},\rep{2})_{7/6}, T_3(\rep{8},\rep{2})_{-1/2}$,\\
& & & & $T_4(\rep{8},\rep{2})_{1/2}, T_5(\rep{1},\rep{2})_{-1/2}, T_6(\rep{1},\rep{2})_{1/2},$\\
& & & & $T_7(\repb{3},\rep{2})_{-7/6}, T_8(\repb{3},\rep{2})_{-1/6}$\\
\hline
Vectors &$\rep{45}$ & $(\rep{6},\rep{2},\rep{2})$ & $(\rep{15},\rep{1},\rep{1}),$ & $(\rep{3},\rep{1})_{2/3}$, $(\repb{3},\rep{1})_{-1/3}$, $(\rep{1},\rep{1})_0$\\
&&& $(\rep{1},\rep{1},\rep{3})$ & $(\rep{1},\rep{1})_1$, $(\rep{1},\rep{1})_{-1}$ \\
\hline\hline
\end{tabular}
\end{table}

\begin{table}[!htb]
\caption{\label{tab:421}Scalar and vector fields that have masses around $\MGUT$ and $\MI$ in the $\G{421}$ model.}
\centering
\begin{tabular}{l l l l l l}
\hline\hline
& $\SO10$ & $\G{421}$ (fields at $\MGUT$) & $\G{421}$ (fields at $\MI$) & $\G{321}$ (fields at $\MI$) \\
\hline
Scalars & $\rep{10}_H$ & $H_1(\rep{6},\rep{1})_0, H_2(\rep{1},\rep{2})_{1/2}$\\
& $\rep{45}_H$ & $\delta_1(\rep{15},\rep{1})_0,\delta_2(\rep{1},\rep{3})_0,$ \\
& & $\delta_3(\rep{6},\rep{2})_{1/2},\delta_4(\rep{6},\rep{2})_{-1/2},$\\
& & $\delta_5(\rep{1},\rep{1})_{-1}, \delta_6(\rep{1},\rep{1})_{0},$ \\ 
& & $\delta_7(\rep{1},\rep{1})_{1}$\\
& $\repb{126}_H$ & $ \Delta_{R1}(\rep{10},\rep{1})_0,\Delta_{R2}(\rep{10},\rep{1})_1$ & $\Delta_{R}(\rep{10}, \rep{1})_{-1},$ & $R_1(\repb{6},\rep{1})_{4/3}, R_2(\rep{3},\rep{1})_{2/3}, R_3(\rep{1},\rep{1})_0$ \\
& & $T_0(\rep{15},\rep{2})_{1/2},$ & $T(\rep{15},\rep{2})_{-1/2}$ &$T_1(\rep{8},\rep{2})_{1/2}, T_2(\rep{3},\rep{2})_{7/6}, T_3(\repb{3},\rep{2})_{-1/6},$ \\
& & & & $T_4(\rep{1},\rep{2})_{1/2}$ \\
& & $\Delta_L(\repb{10},\rep{3})_0,S(\rep{6},\rep{1})_0,$& & \\
\hline
Vectors & $\rep{45}$ & $(\rep{1},\rep{1})_1$, $(\rep{1},\rep{1})_{-1}$, &$(\rep{15},\rep{1})_0$& $(\rep{3},\rep{1})_{2/3}$, $(\repb{3},\rep{1})_{-2/3}$, $(\rep{1},\rep{1})_0$ \\
&& $(\rep{6},\rep{2})_{1/2}$, $(\rep{6},\rep{2})_{-1/2}$&&\\
\hline\hline
\end{tabular}
\end{table}

\begin{table}[!htb]
\caption{\label{tab:3221}Scalar and vector fields that have masses around $\MGUT$ and $\MI$ in the $\G{3221}$ model.}
\centering
\begin{tabular}{l l l l l l}
\hline\hline
& $\SO10$ & $\G{3221}$ (fields at $\MGUT$) & $\G{3221}$ (fields at $\MI$) & $\G{321}$ (fields at $\MI$) \\
\hline
Scalars & $\rep{10}_H$ & $H_1(\rep{3},\rep{1},\rep{1})_{-2/3}, H_2(\repb{3},\rep{1},\rep{1})_{2/3}$ & $\Phi(\rep{1},\rep{2},\rep{2})_0$ & $\phi(\rep{1},\rep{2})_{-1/2}$\\
& $\rep{45}_H$ & $\delta_1(\rep{3},\rep{2},\rep{2})_{-2/3},\delta_2(\repb{3},\rep{2},\rep{2})_{2/3},$ &  & \\
& & $\delta_3(\rep{8},\rep{1},\rep{1})_0,\delta_4(\rep{3},\rep{1},\rep{1})_{4/3},$\\
& & $\delta_5(\repb{3},\rep{1},\rep{1})_{-4/3},\delta_6(\rep{1},\rep{1},\rep{1})_0,$\\
& & $\delta_7(\rep{1},\rep{3},\rep{1})_0$, $\kappa(\rep{1},\rep{1},\rep{3})_0$\\
& $\repb{126}_H$ & $R_1(\repb{6},\rep{1},\rep{3})_{2/3},R_2(\rep{3},\rep{1},\rep{3})_{-2/3},$ & $\Delta_{R}(\rep{1},\rep{1},\rep{3})_{-2},$ & $R_3(\rep{1},\rep{1})_{0}, R_4(\rep{1},\rep{1})_{-1}, R_5(\rep{1},\rep{1})_{-2}$\\
& & $T_1(\rep{8},\rep{2},\rep{2})_0, T_2(\rep{3},\rep{2},\rep{2})_{4/3},$ & $T(\rep{1},\rep{2},\rep{2})_0$ & $T_4(\rep{1},\rep{2})_{-1/2}, T_5(\rep{1},\rep{2})_{1/2}$\\
& & $T_3(\repb{3},\rep{2},\rep{2})_{-4/3},$\\
& & $L_1(\rep{6},\rep{3},\rep{1})_{-2/3},L_2(\repb{3},\rep{3},\rep{1})_{2/3},$\\
& & $L_3(\rep{1},\rep{3},\rep{1})_{2}, S_1(\rep{3},\rep{1},\rep{1})_{-2/3},$\\
& & $S_2(\repb{3},\rep{1},\rep{1})_{2/3}$\\
\hline
Vectors & $\rep{45}$ & $(\rep{3},\rep{2},\rep{2})_{-2/3}$, $(\repb{3},\rep{2},\rep{2})_{2/3}$,& $(\rep{1},\rep{1},\rep{3})_0$&$(\rep{1},\rep{1})_{-1}$, $(\rep{1},\rep{1})_{0}$, $(\rep{1},\rep{1})_{1}$\\
&& $(\rep{3},\rep{1},\rep{1})_{4/3}$, $(\repb{3},\rep{1},\rep{1})_{-4/3}$&& \\
\hline\hline
\end{tabular}
\end{table}

\begin{table}[!htb]
\caption{\label{tab:3221D}Scalar and vector fields that have masses around $\MGUT$ and $\MI$ in the $\G{3221D}$ model.}
\centering
\begin{tabular}{l l l l l l}
\hline\hline
& $\SO10$ & $\G{3221D}$ (fields at $\MGUT$) & $\G{3221D}$ (fields at $\MI$) & $\G{SM}$ (fields at $\MI$) \\
\hline
Scalars & $\rep{10}_H$ & $H_1(\rep{3},\rep{1},\rep{1})_{-2/3}, H_2(\repb{3},\rep{1},\rep{1})_{2/3}$ & $\Phi(\rep{1},\rep{2},\rep{2})_0$ & $\phi(\rep{1},\rep{2})_{-1/2}$\\
& $\rep{45}_H$ & $\delta_1(\rep{3},\rep{2},\rep{2})_{-2/3},\delta_2(\repb{3},\rep{2},\rep{2})_{2/3},$ &$\kappa(\rep{1},\rep{1},\rep{3})_0,$ & $\kappa_1(\rep{1},\rep{1})_{-1},\kappa_2(\rep{1},\rep{1})_{0},\kappa_3(\rep{1},\rep{1})_{1}$\\
& & $\delta_3(\rep{8},\rep{1},\rep{1})_0,$\\
& & $\delta_4(\rep{3},\rep{1},\rep{1})_{4/3},\delta_5(\repb{3},\rep{1},\rep{1})_{-4/3},$ & $\delta_7(\rep{1},\rep{3},\rep{1})_0$ & $\delta_7(\rep{1},\rep{3})_0$\\
& & $\delta_6(\rep{1},\rep{1},\rep{1})_0$\\
& $\repb{126}_H$  & $L_1(\rep{6},\rep{3},\rep{1})_{-2/3},L_2(\repb{3},\rep{3},\rep{1})_{2/3},$ & $\Delta_L(\rep{1},\rep{3},\rep{1})_{2},$ & $L_3(\rep{1},\rep{3})_{1}$\\
& & $R_1(\repb{6},\rep{1},\rep{3})_{2/3},R_2(\rep{3},\rep{1},\rep{3})_{-2/3},$ & $\Delta_{R}(\rep{1},\rep{1},\rep{3})_{-2},$ & $R_3(\rep{1},\rep{1})_{0}, R_4(\rep{1},\rep{1})_{-1}, R_5(\rep{1},\rep{1})_{-2}$\\
& & $T_1(\rep{8},\rep{2},\rep{2})_0, T_2(\rep{3},\rep{2},\rep{2})_{4/3},$ & $T(\rep{1},\rep{2},\rep{2})_0$ & $T_4(\rep{1},\rep{2})_{-1/2}, T_5(\rep{1},\rep{2})_{1/2}$\\
& & $T_3(\repb{3},\rep{2},\rep{2})_{-4/3},$\\
& & $S_1(\rep{3},\rep{1},\rep{1})_{-2/3}, S_2(\repb{3},\rep{1},\rep{1})_{2/3}$\\
& $\rep{210}_H$ & $\Sigma_{L_1}(\rep{8},\rep{3},\rep{1})_0$, $\Sigma_{L_2}(\rep{3},\rep{3},\rep{1})_{4/3}$, & &\\
& &  $\Sigma_{L_3}(\repb{3},\rep{3},\rep{1})_{-4/3}$, $\Sigma_{L_4}(\rep{1},\rep{3},\rep{1})_{0}$, & &\\
& &  $\Sigma_{R_1}(\rep{8},\rep{1},\rep{3})_{0}$, $\Sigma_{R_2}(\rep{3},\rep{1},\rep{3})_{4/3}$, & &\\
& &  $\Sigma_{R_3}(\repb{3},\rep{1},\rep{3})_{-4/3}$, $\Sigma_{R_4}(\rep{1},\rep{1},\rep{3})_{0}$, & &\\
& &  $\xi_{1,1}(\repb{6},\rep{2},\rep{2})_{2/3}$, $\xi_{1,2}(\rep{3},\rep{2},\rep{2})_{-2/3}$, & &\\
& &  $\xi_{1,3}(\rep{1},\rep{2},\rep{2})_{-2}$, $\xi_{2,1}(\rep{6},\rep{2},\rep{2})_{-2/3}$, & &\\
& &  $\xi_{2,2}(\repb{3},\rep{2},\rep{2})_{2/3}$, $\xi_{2,3}(\rep{1},\rep{2},\rep{2})_{2}$, & &\\
& &  $\xi_{3,1}(\rep{8},\rep{1},\rep{1})_{0}$, $\xi_{3,2}(\rep{3},\rep{1},\rep{1})_{4/3}$, & &\\
& &  $\xi_{3,3}(\repb{3},\rep{1},\rep{1})_{-4/3}$, $\xi_{3,4}(\rep{1},\rep{1},\rep{1})_{0}$, & &\\
& &  $\xi_{4,1}(\rep{3},\rep{2},\rep{2})_{-2/3}$, $\xi_{4,2}(\repb{3},\rep{2},\rep{2})_{2/3}$, & &\\
& &  $S^\prime(\rep{1},\rep{1},\rep{1})_{0}$& &\\
\hline
Vectors & $\rep{45}$ & $(\rep{3},\rep{2},\rep{2})_{-2/3}$, $(\repb{3},\rep{2},\rep{2})_{2/3}$,& $(\rep{1},\rep{1},\rep{3})_0$&$(\rep{1},\rep{1})_{-1}$, $(\rep{1},\rep{1})_{0}$, $(\rep{1},\rep{1})_{1}$\\
&& $(\rep{3},\rep{1},\rep{1})_{4/3}$, $(\repb{3},\rep{1},\rep{1})_{-4/3}$&& \\
\hline\hline
\end{tabular}
\end{table}

\clearpage

\section{Threshold Effects}\label{app:thresholds}
In this appendix, we list the threshold corrections for the six models considered. They have been computed using Eq.~\eqref{eq:thresholds} and the table of fields at each scale in App.~\ref{app:fields}. We employ the notation that $\eta_i = \ln (M_i/M)$, where $M_i$ is the mass of each scalar and $M$ is the symmetry breaking scale at which the thresholds apply.

\subsection{Standard Model}\label{app:thresholds_SM}
In the $\G{321}$ model, the threshold corrections at $\MGUT$ are
\begin{align}
\lambda^{10}_3 &= 5 + \eta_{H_1} + \eta_{H_2} + 6 \eta_{A_1} + 6 \eta_{A_2} + 5 \eta_{A_3} + 
 5 \eta_{A_4} + \eta_{A_5} + \eta_{A_6} + \eta_{A_7} + \eta_{A_8} + 
 12 \eta_{B_1} \nonumber \\
 &+ 10 \eta_{B_2} + 2 \eta_{B_3} + 2 \eta_{B_4} + 3 \eta_{C_1} + 
 3 \eta_{C_2} + 2 \eta_{D_1} + 2 \eta_{D_2} + 
 2 \eta_{D_3} + \eta_{E_1} + \eta_{E_2} + 2 \eta_{F_1} \nonumber \\
 &+  3 \eta_{L_2} + 15 \eta_{L_3} + 
  \eta_{R_1} +  \eta_{R_2} + 
  \eta_{R_3} + 5\eta_{R_4} +5 \eta_{R_5} + 5\eta_{R_6} + \eta_{S_1} + \eta_{S_2} + 
2 \eta_{T_1} + 2 \eta_{T_2} \nonumber\\
 &+ 18 \eta_{T_3} + 18 \eta_{T_4} + 2 \eta_{T_7} + 
 2 \eta_{T_8}, \\
\lambda^{10}_2 &= 6 + 8 \eta_{B_1} + 6 \eta_{B_2} + 3 \eta_{B_3} + 3 \eta_{B_4} + 12 \eta_{C_1} + 
 12 \eta_{C_2} + 4 \eta_{C_3} + 4 \eta_{C_4} + 3 \eta_{D_1} + 3 \eta_{D_2} \nonumber\\
&+ 3 \eta_{D_3} + \eta_{D_4} + \eta_{D_5} + \eta_{D_6} + 
 3 \eta_{F_1} + \eta_{F_2} +  4\eta_{L_1} + 12 \eta_{L_2} +  24 \eta_{L_3} + 
 3 \eta_{T_1} + 3 \eta_{T_2} + 8 \eta_{T_3} \nonumber\\
 &+ 8 \eta_{T_4} +  \eta_{T_5} + 
  \eta_{T_6} +1 \eta_{T_7} + 1\eta_{T_8} + \eta_\phi, \\
\lambda^{10}_1 &= 8 + \frac{2}{5}\eta_{H_1} + \frac{2}{5}\eta_{H_2} + \frac{48}{5}\eta_{A_1} + \frac{4}{5}\eta_{A_3} + \frac{16}{5}\eta_{A_4} + \frac{2}{5}\eta_{A_5} + \frac{8}{5}\eta_{A_6} + \frac{2}{5}\eta_{A_7} + \frac{32}{5}\eta_{A_8} \nonumber \\
&+ \frac{24}{5}\eta_{B_1} + \frac{2}{5}\eta_{B_2} + 
 5 \eta_{B_3} + \frac{1}{5}\eta_{B_4} + \frac{6}{5}\eta_{C_1} + \frac{24}{5}\eta_{C_2} + \frac{18}{5}\eta_{C_3} + \frac{49}{5}\eta_{D_1} + \frac{1}{5}\eta_{D_2} + 5 \eta_{D_3} \nonumber\\
 &+ \frac{3}{5}\eta_{D_4} + \frac{3}{5}\eta_{D_5} + \frac{27}{5}\eta_{D_6} + \frac{2}{5}\eta_{E_1} + \frac{8}{5}\eta_{E_2} + \frac{6}{5}\eta_{E_3} + \frac{1}{5}\eta_{F_1} + \frac{3}{5}\eta_{F_2} +  \frac{18}{5}\eta_{L_1} + \frac{6}{5}\eta_{L_2}\nonumber\\
 & + \frac{12}{5}\eta_{L_3} + \frac{8}{5}\eta_{R_1} + \frac{2}{5}\eta_{R_2} + \frac{32}{5}\eta_{R_3} + \frac{64}{5}\eta_{R_4} + \frac{4}{5}\eta_{R_5} + \frac{16}{5}\eta_{R_6} + \frac{24}{5}\eta_{R_7} + \frac{6}{5}\eta_{R_8}  + \frac{2}{5}\eta_{S_1}  \nonumber\\
 &+ \frac{2}{5}\eta_{S_2}+ \frac{1}{5}\eta_{T_1} + \frac{49}{5}\eta_{T_2} + \frac{24}{5}\eta_{T_3} + \frac{24}{5}\eta_{T_4} + \frac{3}{5}\eta_{T_5} + \frac{3}{5}\eta_{T_6} + \frac{49}{5}\eta_{T_7} + \frac{1}{5}\eta_{T_8}+ \frac{3}{5}\eta_{\phi}.
\end{align}

\subsection{$\SU\text{(4)}\times\SU\text{(2)}\times\SU\text{(2)}$}
In the $\G{422}$ model, the threshold corrections at $\MGUT$ are
\begin{align}
\lambda^{10}_4 &= 4 + 2\eta_H + 2\eta_S + 18\eta_{\Delta_L} + 12\eta_{\Sigma_L} + 12\eta_{\Sigma_R} + 12\eta_{\xi_1} + 12\eta_{\xi_2} + 4\eta_{\xi_3} + 4\eta_{\xi_4} + 8\eta_{\delta_1} \nonumber\\
&+ 8\eta_{\delta_3}, \\
\lambda^{10}_\text{2L} &= 6 + 40\eta_{\Delta_L} + 30\eta_{\Sigma_L} + 10\eta_{\xi_1} + 10\eta_{\xi_2} + 6\eta_{\xi_4} + 4\eta_{\delta_2} + 12\eta_{\delta_3}, \\
\lambda^{10}_\text{2R} &= 6 + 30\eta_{\Sigma_R} + 10\eta_{\xi_1} + 10\eta_{\xi_2} + 6\eta_{\xi_4} + 12 \eta_{\delta_3}.
\end{align}
At $\MI$, they are
\begin{align}
\lambda^{422}_3 &= 1 + \eta_{R_1} + \eta_{R_2} + \eta_{R_3} + 5\eta_{R_4} + 5\eta_{R_5} + 5\eta_{R_6}+ 2\eta_{T_1} + 2\eta_{T_2} + 12\eta_{T_3} + 12\eta_{T_4} + 2\eta_{T_7}  \nonumber\\
&+ 2\eta_{T_8}, \\
\lambda^{422}_2 &= \eta_\phi + 3\eta_{T_1} + 3\eta_{T_2} + 8\eta_{T_3} + 8\eta_{T_4} + \eta_{T_5} + \eta_{T_6} + 3\eta_{T_7} + 3\eta_{T_8} , \\
\lambda^{422}_1 &= \frac{14}{5} + \frac{3}{5}\eta_\phi + \frac{6}{5}\eta_{\kappa_1} + \frac{6}{5}\eta_{\kappa_2} + \frac{2}{5}\eta_{R_1} + \frac{32}{5}\eta_{R_2} + \frac{8}{5}\eta_{R_3} + \frac{64}{5}\eta_{R_4} + \frac{4}{5}\eta_{R_5} + \frac{16}{5}\eta_{R_6} \nonumber\\
&+ \frac{24}{5}\eta_{R_7} + \frac{6}{5}\eta_{R_8} + \frac{1}{5}\eta_{T_1} + \frac{49}{5}\eta_{T_2} + \frac{24}{5}\eta_{T_3} + \frac{24}{5}\eta_{T_4} + \frac{3}{5}\eta_{T_5} + \frac{3}{5}\eta_{T_6} + \frac{49}{5}\eta_{T_7} + \frac{1}{5}\eta_{T_8}.
\end{align}

\subsection{$\SU\text{(4)}\times\SU\text{(2)}\times\SU\text{(2)}\times \text{D}$}
In the $\G{422D}$ model, the threshold corrections at $\MGUT$ are
\begin{align}
\lambda_4^{10} &= 4 + 2\eta_{S} + 2\eta_{H} + 8\eta_{\delta_1} + 8\eta_{\delta_3} + 8\eta_{\zeta_2} + 16\eta_{\zeta_3}, \\
\lambda_\text{2L}^{10} &= 6 + 12\eta_{\delta_3} + 12\eta_{\zeta_1} + 12\eta_{\zeta_2},\\
\lambda_\text{2R}^{10} &= 6 + 12\eta_{\delta_3} + 12\eta_{\zeta_1} + 12\eta_{\zeta_2}.
\end{align}
At $\MI$, they are
\begin{align}
\lambda_3^\text{422D} &= 1 + 3\eta_{L_2} + 15\eta_{L_3} + \eta_{R_1} + \eta_{R_2} + \eta_{R_3} + 5 \eta_{R_4} + 5 \eta_{R_5} + 5\eta_{R_6} + 2\eta_{T_1} + 2\eta_{T_2} \nonumber\\
&+ 12\eta_{T_3} + 12\eta_{T_4} + 2\eta_{T_7} + 2\eta_{T_8},\\
\lambda_2^\text{422D} &= 4\eta_{L_1} + 12\eta_{L_2} + 24\eta_{L_3} + 3\eta_{T_1} + 3\eta_{T_2} + 8\eta_{T_3} + 8\eta_{T_4} + \eta_{T_5} + \eta_{T_6} + 3\eta_{T_7} \nonumber\\
&+ 3\eta_{T_8} + 4\eta_{\delta_2} + \eta_\phi, \\
\lambda_1^\text{422D} &= \frac{14}{5} + \frac{6}{5} \eta_{\kappa_1}+\frac{6}{5}\eta_{\kappa_2}+\frac{6}{5}\eta_{L_1}+\frac{2}{5}\eta_{L_2}+\frac{4}{5} \eta_{L_3}+\frac{2}{5}\eta_{R_1}+\frac{32}{5}\eta_{R_2}+\frac{8}{5} \eta_{R_3}+\frac{64}{5}\eta_{R_4}\nonumber\\
   &+\frac{4}{5}\eta_{R_5}+\frac{16}{5}\eta_{R_6}+\frac{24}{5}\eta_{R_7}+\frac{6}{5}\eta_{R_8}+\frac{1}{5}\eta_{T_1}+\frac{49}{5} \eta_{T_2}+\frac{24}{5}\eta_{T_3}+\frac{24}{5}\eta_{T_4}+\frac{3}{5}\eta_{T_5}\nonumber\\
   &+\frac{3}{5}\eta_{T_6}+\frac{49}{5}\eta_{T_7}+\frac{1}{5}\eta_{T_8}+\frac{3}{5} \eta_\phi.
\end{align}

\subsection{$\SU\text{(4)}\times\SU\text{(2)}\times\text{U(1)}$}
In the $\G{421}$ model, the threshold corrections at $\MGUT$ are
\begin{align}
\lambda^{10}_4 &= 4 + 2\eta_{H_1} + 2\eta_S + 16\eta_{T} + 18\eta_{\Delta_L} + 8\eta_{\delta_1} + 4\eta_{\delta_3} + 4\eta_{\delta_4} + 6\eta_{\Delta_{R1}} + 6\eta_{\Delta_{R2}}, \\
\lambda^{10}_2 &= 6 + \eta_{H_2} + 15\eta_{T} +40\eta_{\Delta_L} + 4\eta_{\delta_2} + 6\eta_{\delta_3} + 6\eta_{\delta_4}, \\
\lambda^{10}_1 &= 8 + \eta_{H_2} + 20\eta_{R_2} + 15\eta_{T} + 6\eta_{\delta_3} + 6\eta_{\delta_4} + 2\eta_{\delta_5} + 2\eta_{\delta_7}.
\end{align}
At $\MI$, they are
\begin{align}
\lambda^{421}_3 &= 1 + 5\eta_{R_1} + \eta_{R_2} + 12 \eta_{T_1} + 2\eta_{T_2} + 2\eta_{T_3}, \\
\lambda^{421}_2 &= 8\eta_{T_1} + 3\eta_{T_2} + 3\eta_{T_3} + \eta_{T_4}, \\
\lambda^{421}_1 &= \frac{8}{5} + \frac{64}{5}\eta_{R_1} + \frac{8}{5} \eta_{R_2} + \frac{12}{5}\eta_{T_1} + \frac{49}{5}\eta_{T_2} + \frac{1}{5}\eta_{T_3} + \frac{3}{5}\eta_{T_4}.
\end{align}

\subsection{$\SU\text{(3)}\times\SU\text{(2)}\times\SU\text{(2)}\times \text{U(1)}$}
In the $\G{3221}$ model, the threshold corrections at $\MGUT$ are
\begin{align}
\lambda^{10}_3 &= 5 + \eta_{H_1} + \eta_{H_2} + 15 \eta_{L_1} + 3 \eta_{L_2} + 15 \eta_{R_1} + 
 3 \eta_{R_2} + \eta_{S_1} + \eta_{S_2} + 24 \eta_{T_1} + 4 \eta_{T_2} \nonumber\\
 &+  4 \eta_{T_3} + 4 \eta_{\delta_1} + 4 \eta_{\delta_2} + 
 6 \eta_{\delta_3} + \eta_{\delta_4} + \eta_{\delta_5}, \\
\lambda^{10}_\text{2L} &= 6 + 24 \eta_{L_1} + 12 \eta_{L_2} + 4 \eta_{L_3} + 16 \eta_{T_1} + 
 6 \eta_{T_2} + 6 \eta_{T_3} + 6 \eta_{\delta_1} + 6 \eta_{\delta_2} + 
 4 \eta_{\delta_7}, \\
\lambda^{10}_\text{2R} &= 6 + 24 \eta_{R_1} + 12 \eta_{R_2} + 16 \eta_{T_1} + 6 \eta_{T_2} + 
 6 \eta_{T_3} + 6 \eta_{\delta_1} + 6 \eta_{\delta_2} + 4 \eta_\kappa, \\
\lambda^{10}_{1} &= 8 + \eta_{H_1} + \eta_{H_2} + 6 \eta_{L_1} + 3 \eta_{L_2} + 9 \eta_{L_3} + 
 6 \eta_{R_1} + 3 \eta_{R_2} + \eta_{S_1} + \eta_{S_2} + 16 \eta_{T_2} \nonumber\\
 &+ 16 \eta_{T_3} + 4 \eta_{\delta_1} + 4 \eta_{\delta_2} + 
 4 \eta_{\delta_4} + 4 \eta_{\delta_5}.
\end{align}
At $\MI$, they are
\begin{align}
\lambda^{3221}_3 &= 0, \\
\lambda^{3221}_2 &= \eta_{T_4} + \eta_{T_5} + \eta_{\phi}, \\
\lambda^{3221}_1 &= \frac{6}{5} + \frac{6}{5}\eta_{R_4} + \frac{24}{5}\eta_{R_5} + \frac{3}{5}\eta_{T_4} + \frac{3}{5}\eta_{T_5} + \frac{3}{5}\eta_{\phi}.
\end{align}

\subsection{$\SU\text{(3)}\times\SU\text{(2)}\times\SU\text{(2)}\times \text{U(1)}\times \text{D}$}
In the $\G{3221D}$ model, the threshold corrections at $\MGUT$ are
\begin{align}
\lambda^{10}_3 &= 5 + \eta_{H_1} + \eta_{H_2} + 15 \eta_{L_1} + 3 \eta_{L_2} + 15 \eta_{R_1} + 
 3 \eta_{R_2} + \eta_{S_1} + \eta_{S_2} + 24 \eta_{T_1} + 4 \eta_{T_2} + 
 4 \eta_{T_3} \nonumber\\
 &+ 4 \eta_{\delta_1} + 4 \eta_{\delta_2} + 
 6 \eta_{\delta_3} + \eta_{\delta_4} + \eta_{\delta_5} + 
 20 \eta_{\xi_{1_1}} + 4 \eta_{\xi_{1_2}} + 20 \eta_{\xi_{2_1}} + 
 4 \eta_{\xi_{2_2}} + 6 \eta_{\xi_{3_1}} + \eta_{\xi_{3_2}} \nonumber\\
 & + \eta_{\xi_{3_3}} + 4 \eta_{\xi_{4_1}} + 4 \eta_{\xi_{4_2}} + 18 \eta_{\Sigma_{L_1}} + 
 3 \eta_{\Sigma_{L_2}} + 3 \eta_{\Sigma_{L_3}} + 
 18 \eta_{\Sigma_{R_1}} + 3 \eta_{\Sigma_{R_2}} + 
 3 \eta_{\Sigma_{R_3}}, \\
\lambda^{10}_\text{2L} &= 6 + 24 \eta_{L_1} + 12 \eta_{L_2} + 16 \eta_{T_1} + 6 \eta_{T_2} + 
 6 \eta_{T_3} + 6 \eta_{\delta_1} + 6 \eta_{\delta_2} + 
 12 \eta_{\xi_{1_1}} + 6 \eta_{\xi_{1_2}} + 2 \eta_{\xi_{1_3}} \nonumber\\
 &+ 12 \eta_{\xi_{2_1}} + 6 \eta_{\xi_{2_2}} + 2 \eta_{\xi_{2_3}} + 
 6 \eta_{\xi_{4_1}} + 6 \eta_{\xi_{4_2}} + 32 \eta_{\Sigma_{L_1}} + 
 12 \eta_{\Sigma_{L_2}} + 12 \eta_{\Sigma_{L_3}} + 
 4 \eta_{\Sigma_{L_4}}, \\
\lambda^{10}_\text{2R} &= 6 + 24 \eta_{R_1} + 12 \eta_{R_2} + 16 \eta_{T_1} + 6 \eta_{T_2} + 
 6 \eta_{T_3} + 6 \eta_{\delta_1} + 6 \eta_{\delta_2} + 
 12 \eta_{\xi_{1_1}} + 6 \eta_{\xi_{1_2}} + 2 \eta_{\xi_{1_3}} \nonumber\\
 &+ 12 \eta_{\xi_{2_1}} + 6 \eta_{\xi_{2_2}} + 2 \eta_{\xi_{2_3}} + 
 6 \eta_{\xi_{4_1}} + 6 \eta_{\xi_{4_2}} + 36 \eta_{\Sigma_{L_3}} + 
 12 \eta_{\Sigma_{L_4}} + 32 \eta_{\Sigma_{R_1}} + 
 12 \eta_{\Sigma_{R_2}}, \\
\lambda^{10}_1 &= 8 + \eta_{H_1} + \eta_{H_2} + 6 \eta_{L_1} + 3 \eta_{L_2} + 6 \eta_{R_1} + 
 3 \eta_{R_2} + \eta_{S_1} + \eta_{S_2} + 16 \eta_{T_2} + 16 \eta_{T_3} + 4 \eta_{\delta_{1}} \nonumber\\
 &+ 4 \eta_{\delta_{2}} + 4 \eta_{\delta_{4}} + 
 4 \eta_{\delta_{5}} + 8 \eta_{\xi_{1_1}} + 4 \eta_{\xi_{1_2}} + 
 12 \eta_{\xi_{1_3}} + 8 \eta_{\xi_{2_1}} + 4 \eta_{\xi_{2_2}} + 
 12 \eta_{\xi_{2_3}} + 4 \eta_{\xi_{3_2}} \nonumber\\
 &+ 4 \eta_{\xi_{3_3}} +  4 \eta_{\xi_{4_1}} + 4 \eta_{\xi_{4_2}} + 12 \eta_{\Sigma_{L_2}} + 
 12 \eta_{\Sigma_{L_3}} + 12 \eta_{\Sigma_{R_2}} + 
 12 \eta_{\Sigma_{R_3}}.
\end{align}
At $\MI$, they are
\begin{align}
\lambda^\text{3221D}_3 &= 0, \\
\lambda^\text{3221D}_2 &= 4 \eta_{L_3} + \eta_{T_4} + \eta_{T_5} + 4 \eta_{\delta_7} + \eta_{\phi}, \\
\lambda^\text{3221D}_1 &= \frac{6}{5} + \frac{18}{5}\eta_{L_3} + \frac{6}{5}\eta_{R_4} + \frac{24}{5}\eta_{R_5} + \frac{3}{5}\eta_{T_4} + \frac{3}{5}\eta_{T_5} + \frac{6}{5}\eta_{\kappa_1} + \frac{6}{5}\eta_{\kappa_2} + \frac{3}{5}\eta_{\phi}.
\end{align}

\begin{acknowledgments}
The authors wish to thank Sofiane M.~Boucenna for collaboration in early stages of this project and Erik S{\"o}nnerlind for pointing out errors and clarifying issues with numerics in a draft of this manuscript. T.O.~acknowledges support by the Swedish Research Council (Vetenskapsrådet) through contract No.~2017-03934 and the KTH Royal Institute of Technology for a sabbatical period at the University of Iceland. M.P.~thanks “Stiftelsen Olle Engkvist Byggmästare” and “Roland Gustafssons Stiftelse för teoretisk fysik” for financial support. Numerical computations were performed on resources provided by the Swedish National Infrastructure for Computing (SNIC) at PDC Center for High Performance Computing (PDC-HPC) at KTH Royal Institute of Technology in Stockholm, Sweden under project numbers SNIC 2018/3-559 and SNIC 2020/5-122.
\end{acknowledgments}

\bibliographystyle{apsrev4-1}
\bibliography{refs_thresholds.bib}

\end{document}